\documentclass[fleqn,usenatbib]{mnras}

\usepackage{newtxtext,newtxmath}
\usepackage[T1]{fontenc}

\DeclareRobustCommand{\VAN}[3]{#2}
\let\VANthebibliography\thebibliography
\def\thebibliography{\DeclareRobustCommand{\VAN}[3]{##3}\VANthebibliography}

\usepackage{graphicx}	% Including figure files
\usepackage{xspace}

\newcommand{\pc}{\,\mathrm{pc}}
\newcommand{\Msun}{\,\mathrm{M}_{\odot}}

\newcommand{\Myr}{\,\mathrm{Myr}}
\newcommand{\Myrs}{\xspace\mathrm{Myrs}}

\newcommand{\K}{\,\mathrm{K}}

\newcommand{\kms}{\,\mathrm{km\,s}^{-1}}
\newcommand{\gocm}{\,\mathrm{g\,cm}^{-3}}

\newcommand{\MLL}{{\sc M5Infall24}\xspace}
\newcommand{\MLH}{{\sc M5Infall23}\xspace}

\newcommand{\MHL}{{\sc M6Infall24}\xspace}
\newcommand{\MHH}{{\sc M6Infall23}\xspace}

\newcommand{\NONRT}{{\sc NoRT}\xspace}
\newcommand{\RT}{{\sc RT}\xspace}

\newcommand{\ramses}{{\sc Ramses}\xspace}
\newcommand{\ramsesrt}{{\sc Ramses-rt}\xspace}

\newcommand{\tinf}{t_{\rm I}}
\newcommand{\tAGB}{t_{\rm AGB}}

\defcitealias{calura19}{C19}
\defcitealias{Yaghoobi2022a}{PaperI}
\defcitealias{Yaghoobi2022b}{PaperII}

\newcommand{\secref}[1]{Section~\ref{#1}}
\newcommand{\figref}[1]{Figure~\ref{#1}}
\newcommand{\tabref}[1]{Table~\ref{#1}}

\usepackage[normalem]{ulem}
\usepackage{xcolor}
\title[Photoionization feedback on SG star formation]{On the effects of photoionization feedback on second-generation star formation in globular clusters of different masses}

\author[A. Yaghoobi et al.]{
A. Yaghoobi,$^{1,2}$\thanks{E-mail:asiyeh.yaghoobi@gmail.com}
J. Rosdahl$^{2}$,
F. Calura$^{3}$ and  
S. Ataiee$^{1}$
\\
$^{1}$ Department of Physics, Faculty of Sciences, Ferdowsi University of Mashhad, Mashhad, 91775-1436, Iran \\
$^{2}$ Univ Lyon, Univ Lyon1, Ens de Lyon, CNRS, Centre de Recherche Astrophysique de Lyon UMR5574, F-69230, Saint-Genis-Laval, France\\
$^{3}$ INAF - OAS, Osservatorio di Astrofisica e Scienza dello Spazio di Bologna, via Gobetti 93/3, I-40129 Bologna, Italy
}

\date{Accepted XXX. Received YYY; in original form ZZZ}

\pubyear{2015}

\begin{document}
\label{firstpage}
\pagerange{\pageref{firstpage}--\pageref{lastpage}}
\maketitle

\begin{abstract}
We simulate the formation of second-generation stars in young clusters with masses of $10^5$ and $10^6\Msun$ within $30-100\Myr$ after the formation of clusters. We assume the clusters move through a uniform interstellar medium with gas densities of $10^{-24}$ and $10^{-23}\gocm$ and consider the stellar winds from asymptotic giant branch (AGB) stars, gas accretion onto the cluster, ram pressure, star formation, and photoionization feedback of our stellar systems including binary stars.
We find that second generation (SG) stars can be formed only within the $10^6\Msun$ cluster in the high-density simulation, where the cluster can accrete sufficient pristine gas from their surrounding medium, leading to efficient cooling required for the ignition of SG formation and sufficient dilution of the AGB ejecta. 
Hence, our results indicate that a denser environment is another requirement for the AGB scenario to explain 
the presence of multiple populations in globular clusters. 
On the other hand, the ionizing feedback becomes effective in heating the gas in our low-density simulations. As a result, the clusters cannot accumulate a considerable amount of pristine gas at their center. The gas mass within the clusters in these simulations is similar to that in young massive clusters (YMCs). Hence, our studies can provide a possible reason for the lack of gas, star formation, and SG stars in YMCs. Our results indicate that the ionizing stellar feedback is not a severe problem for SG formation; rather, it can help the AGB scenario to account for some observables.
\end{abstract}


%%%%%%%%%%%%%%%%%%%%%%%%%%%%%%%%%%%%%%%%%%%%%%%%%%

%%%%%%%%%%%%%%%%% BODY OF PAPER %%%%%%%%%%%%%%%%%%

\section{Introduction}
Globular clusters (GCs), containing low-metallicity ([Fe/H]$\lesssim 1$ ), low-mass ($\lesssim 0.75 \Msun$), and old ($\gtrsim 1 {\rm Gyr}$) stars  \citep{cadelano20}, can provide useful information as early star-forming subsystems to study the formation history of their parent galaxies \citep{Larson1996}. Recent observations have revealed that GC stars display anomalous variations in light elements (e. g. He, C, N, O, Na, Al, and sometimes Mg) that are not expected to be due low-mass stars during stellar evolutionary processes \citep{Prantzos2007,carretta2009, Piotto2015}. Together with another population characterized  by a composition similar to normal field stars, such \textit{anomalous} stars build Multiple Stellar Populations (MSPs) observed in GCs. 
The anomalous stars are enriched in N and Na and depleted in C and O, whereas the sum of the C, N, and O abundances inside these stars are observed to be generally constant within the measurement errors \citep{Dickens1991,Yong2015}.
Additionally, except for some massive GCs such as $\omega$~Cen \citep{Johnson2010}, no considerable spread in Fe abundance and heavy elements has been found in GCs hosting MSPs \citep{Gratton2004,Gratton2012}. This specific abundance pattern makes it difficult to explain how MSPs have been formed in GCs; hence their formation remains a mystery. 

Such an abundance pattern is expected from the yields of hot hydrogen burning within intermediate-mass and massive stars \citep{renzini2015}. Hence, most scenarios proposed for the formation of MSPs are based on the ejecta of these stars, fuelling the formation of the SG stars, where MSPs are subsequent generations. The population with the same composition as the field stars at the same metallicity [Fe/H] is referred to as the first generation (FG) or first population. Stars within GCs showing variation in the light elements compared to the first population are commonly categorized as the second generation (SG) or second population \citep{bastian2018,milone2022}. To date, the suggested FG candidates  are intermediate-mass asymptotic giant branch stars (AGBs; $4\leq m/\Msun\leq 8$; \citealt{D'Ercole2008,Conroy2011,D'Ercole2016,Bekki2017}), fast rotating massive stars (FRMS, $20\leq m/\Msun\leq120$; \citealt{decressin2007,DecCharbMey2007,krause2013}), and supermassive stars ($m/\Msun\geq10^4$; \citealt{Denissenkov2014,Denissenkov2015}).
Alternative scenarios propose different mechanisms in which the MSPs may be coeval. For example, in the early disc accretion scenario, the FG low-mass stars accrete enriched stellar winds from more massive interacting binary stars, resulting in different abundances \citep{bastian2013Earlydisc}. 

Each proposed scenario has its own advantages and drawbacks, but none of them has been able to explain all the observed properties of MSPs
\cite[see][]{renzini2015,bastian2018}, such as the observed chemical anomalies. 
To solve this issue, the ejecta need to be diluted by pristine gas, i.e. the gas with the same composition as the one that formed the FG stars. In addition, most models, such as AGB and FRMS scenarios, encounter a "mass budget problem" to produce a cluster with a SG fraction similar to observed values, between $\sim 30$ and $90$ percent for Galactic GCs \citep{milone2017}. One problem is that, assuming a standard initial mass function (IMF),  the FG ejecta do not contain enough mass to make up the entirety of the observed SG mass. This problem might be alleviated in specific conditions, such as a top-heavy IMF for FG stars and a binary fraction close to unity \citep{D'Ercole2008,Khalaj2015,Krause2016,Vesperini2021}. While generally not clear from observations, the epoch of SG formation and age difference between populations depend on the proposed scenario. In models based on massive and very massive stars, SG formation should occur within a short time-scale after the formation of FG stars and before the start of type II supernova explosions to avoid the incorporation of newly produced Fe and a significant metallicity spread \citep[e. g.][]{krause2013}. On the other hand, in the AGB scenario, the SG formation starts after the end of supernovae (SNe) (about $30\Myr$) and lasts for roughly a few ten Myrs.  \citep{D'Ercole2008,D'Ercole2016,calura19}. 

The presence of MSPs with anomalous chemical compositions is found to be specific to only massive GCs older than $2\, \mathrm{Gyr}$ \citep{bastian2018}. It is not clear yet whether young massive clusters (YMCs, with a mass range of $[10^4 - 10^{5.5}]\Msun$ and ages $\lesssim 100\Myr$) host MSPs with such anomalous chemical compositions \citep{Portegies2010}. 
Nevertheless, observationally these clusters do not show any significant amount of gas \citep{Bastian2013I,Bastian2014IV,Hollyhead2015,Cabrera2015V} and any evidence for ongoing star formation (SF) \citep{Bastian2013I}. As their age can be compared with the proposed epoch for SG formation in scenarios, in principle  these facts can constrain SG formation models \citep{Bastian2013I,Bastian2014III,Bastian2014IV}, assuming that YMCs are equivalent to proto-GCs.
But some rely on the idea that the SG formation is limited to only the early Universe, not the present-day one \citep{D'Ercole2016,renzini2015}. For example, \cite{D'Ercole2016} show that the AGB scenario needs ambient surrounding properties of GCs consistent with those of star-forming disk galaxies observed at redshifts $> 2$. Moreover, the fact that YMCs are almost gas free \citep{Bastian2013I,Bastian2014IV,Hollyhead2015,Cabrera2015V} indicates that they efficiently expel their intracluster medium (ICM), despite significant mass-loss from their stars and their capability to accrete gas from the ambient medium. This raises another question: under which physical conditions can a young cluster get rid of its ICM?
To answer this question, we require more comprehensive theoretical studies of the gas dynamics in young clusters, including the key processes, including supernova (SN) feedback, ram pressure, radiative feedback, cluster potential, and stellar wind heating \citep{Conroy2011,calura2015,gavagnin2017, Wunsch2017, Naiman2018, Chantereau2020}.  

Since a few years, we have focused on the gas accumulation and SG formation in young ($>30\Myr$) clusters with different masses $[10^5-10^7 \Msun]$ based on the AGB scenario \citep{calura19,lacchin21,Yaghoobi2022a,Yaghoobi2022b,Lacchin2022}. In this scenario, the origin of SG stars is a mixture of the AGB stellar winds and pristine gas accumulated in clusters.  \cite{D'Ercole2016} investigate the origin of the pristine gas and its dynamics during the phase of SG formation. Their model follows three requirements; i) the pristine gas should initially be removed from the cluster by type II SN explosions within about $30\Myr$ after the birth of the first generation of stars; ii) the cluster should be able to accrete the pristine gas within about $60\Myr$; iii) the SN II ejecta should not contaminate the pristine gas. They show that these conditions can explain the origin of SG stars only in massive clusters that formed in the disks of galaxies at $z > 2 $\citep{Kravtsov2005,Kruijssen2015}.

Using 3D hydrodynamical simulations, we investigated this model for clusters of different masses in \cite{calura19} and \cite{Yaghoobi2022a} (hereafter \citetalias{Yaghoobi2022a}).
We performed a parameter study of the gas density of the  ISM with values $10^{-24}$ and $10^{-23}\gocm$, roughly corresponding to the values observed in star-forming galaxies at low and high redshifts \citep{marcolini2003,wardlow2017, D'Ercole2016} to explore the effect of different environments on the formation of SG stars. We found that the properties of the SG population depend significantly on the surrounding environment of the clusters. In the high-density simulations, massive clusters (with masses $ \geq 10^6\Msun$) could overcome the ram pressure, accrete pristine gas, and retain their own stellar winds, producing a massive SG. Positive correlations were also found for the SG mass fraction and maximum He enhancement in GCs versus the cluster mass.

In \cite{Yaghoobi2022b}, hereafter \citetalias{Yaghoobi2022b}, we included ionizing radiation from stars in our simulations and found that the gravitational potential of a very massive cluster ($10^7\Msun$, with half-mass radius $30 \pc$) is strong enough to accumulate both accreted pristine gas and AGB ejecta at its center.  Photoionization heating was found to be efficient in warming up and expanding the gas for only about the first $50$ Myrs from the FG formation. After that, the gas cooling dominates the radiative heating 
and the gas accumulated in the innermost regions results in star formation. 
However, due to ionizing feedback effects, the total SG mass became a factor two lower than in the run without ionizing radiation at the end of simulations ($100 \Myr$ after the cluster formation). Nevertheless, the SG mass fraction within the cluster reached $40\%$ in the high-density model, comparable with observed fractions for Galactic GCs \citep{milone2017}. However, in the low-density case, the gas content and SG fraction were very low ($\sim 4\%$). To further explore the viability of this scenario, our study including the photoionization feedback needs to be expanded to clusters with different structural parameters, since it is shown that the cooling and heating processes depend on the properties of the cluster \citep{Wunsch2017,Yaghoobi2022a}. Particularly, we are interested in studying the SF in the clusters expected to be equivalent to proto-GCs and YMCs to explore the role of stellar feedback in SG formation. Moreover, we investigate how results are sensitive to the different environments, which are expected to be different for GCs and YMCs. 

This study is a follow-up to \citetalias{Yaghoobi2022b} to investigate the SF and gas content in young clusters with different properties (mass and half-mass radius). We aim at checking whether these clusters can retain their stellar winds, accrete the pristine gas from the ambient medium, and cool their ICM in the presence of ionizing radiation.
The paper is organized as follows. In Section \ref{sec:setup}, we describe the simulation set-up and main assumptions of our model. In Section \ref{sec:results}, we present the results of our simulations, and in Section \ref{sec:discussion} discuss them. Finally, in Section \ref{Conclusions}, we list the main conclusions of our study.

%%%%%%%%%%%%%%%%%%%%%%%%%%%%%%%%%%%%%%%%%%%%%%%%%%%%%%%%%%%%%%%%%%%%%%%%%%%%%%%%%%%%%%%%%%%%%%%%%%%%%%%%%%%
\section{Simulation setup}\label{sec:setup}
The simulation setup in this study is the same as \citetalias{Yaghoobi2022b} for a cluster with the mass of $10^7\Msun$ and half-mass radius of $23\pc$, except that we study the SG formation within FG clusters with masses of $10^5$ and $10^6\Msun$, and half-mass radius of $4\pc$, moving through a homogeneous ISM with uniform density. These masses are comparable with the initial masses derived for GCs by calculating the cluster orbits backward in time \citep{Baumgardt2019}. 
To study the role of environments in SF formation, we consider different values for ISM densities; $10^{-24}\gocm$ for a typical dwarf galaxy \citep{marcolini2003} and $10^{-23}\gocm$ as a representative value for star-forming regions in galaxies at high redshifts \citep{wardlow2017}.
%and  $10^{-23}\gocm$ for a merging system such as the Antennae (Zhu, Seaquist and Kuno 2003 ).
As in \citetalias{Yaghoobi2022b}, we consider the effects of the ionizing FG and SG radiation using radiation-transfer simulations  (\RT simulations).
We also run their analogs without radiation (\NONRT) to better understand the role of stellar radiative feedback in SG formation. The main simulations of this study and their relevant characteristics are listed in \tabref{tabl1}.

\begin{table}
	\centering
	\caption{Main parameters of our simulations for both cases of \RT and \NONRT. Column description: M$_{\rm FG}$ is the mass of the cluster; $\rho_{\rm pg}$ the gas density of the pristine gas;  $\tinf$ the time when the cluster reaches the ISM (\secref{sec:init}) at the time reference of this paper (the FG formation). Note that the AGB ejecta in all simulations start at $t_{\rm AGB}=39\Myr$.}  
	\begin{tabular}{lcccl} 
		\hline
		Simulation & M$_{{\rm FG}}$ \ ${\rm [M_{\odot}]}$  & $\rho_{\rm pg}$ \ $[\rm g \ cm^{-3}]$  & $\tinf$ \ $[\rm Myr]$ \\		\hline
		\MLL  & $10^5$ &  $10^{-24}$ &   33.9 \\
		\MLH & $10^5$  & $10^{-23}$&  30.0 \\
		%\MLVH & $10^5$ &  $10^{-22}$ &  30 &  RT \\
		\MHL  & $10^6$ &  $10^{-24}$ &   42.5 \\
		\MHH & $10^6$  & $10^{-23}$&  33.9 \\
		%\MHVH & $10^6$ &  $10^{-22}$ &  30 &  RT \& NoRT\\
		\hline
	\end{tabular}
	
	\label{tabl1}
\end{table}
We use \ramsesrt \citep{rosdahl2013,Rosdahl2015}, the radiative hydrodynamics version of the \ramses \citep{teyssier2002} code. The code uses a second-order Godunov scheme to solve the Euler equations and a particle-mesh solver to compute the dynamical evolution of particles.
%check ST
In \citetalias{Yaghoobi2022a}, the appropriate simulation volume for the $10^5$  and $10^6\Msun$ clusters was found to be about $50 \pc$. However, we find that we need to consider a larger box to achieve convergence in results due to gas expansion resulting from photoionization heating. We find that a cube with a width of $128\pc$ and inflow boundaries can fulfill this goal. 
Moreover, we use the AMR strategy with a maximum size of $\Delta x_{\rm max}= 1\pc$ ($l_{\rm min}=7$ in the \ramses code) and a minimum size $\Delta x_{\rm min}= 0.062\pc$ ($l_{\rm max} = 11$). Cells are refined to smaller sizes if their mass exceeds $10\Msun$.
As in \citetalias{Yaghoobi2022a}, we performed a series of tests assuming different resolutions, ranging from $\Delta x_{\rm min}= 1\pc$ to $\Delta x_{\rm min}=0.062$, indicating that the chosen setup is sufficient for our results to converge. 
 
The simulation setup of our model has been discussed extensively in \citetalias{Yaghoobi2022b}. Here, we briefly recap it and encourage the reader to consult \citetalias{Yaghoobi2022a}, \citetalias{Yaghoobi2022b}, and \citet{calura19} for more details. In our notation, time $t=0$ corresponds to the birth time of the first generation of stars.

%%%%%%%%%%%%%%%%%%%%%%%%%%%%%%%%%%%%%%%%%%%%%%%%%%%%%%%%%%%%%%%%%%%%%%%%%%%%%%%%%%%%%%%%%%%
\subsection{Initial conditions}\label{sec:init}
We set the initial conditions of our simulations as proposed in \citet{D'Ercole2016}. Type II SN explosions of FG stars have formed a bubble of hot and diffuse gas around the cluster at the end of their activity, assumed to be $30\Myr$ after the birth of the first generation. The bubble radius depends on the cluster mass, ISM density, and velocity with respect to the gas \citep[see][]{calura19}. 
It is assumed that the bubble is filled with extremely diffuse gas in hydrostatic equilibrium with the cluster. Thus, FG stars are not initially in direct contact with the pristine ISM gas until they traverse this radius and reach the ISM. After that, the pristine gas encounters the cluster and starts to dilute the enriched AGB ejecta. We also assume that the injection of AGB ejecta starts at $  t_{\rm AGB} = 39 \Myr$.
Following this model, we consider the velocity of clusters to be $23\kms$, of the order of the gas velocity dispersion in an ionized medium, and assume it to be constant during the simulations. Accordingly, the time at which the cluster reaches the ISM ($ t_{\rm I}$) can be estimated, as described in \citetalias{Yaghoobi2022a}. $t_{\rm I}$s for our simulations are computed by means of Eq. 11  of \citet{D'Ercole2016} and presented in the fourth column of Table \ref{tabl1}. With the exception of the \MHL model,  the infall starts earlier than stellar winds in all our simulations. This is visible from the comparison of $t_{\rm I}$ and $t_{\rm AGB}$ in Table \ref{tabl1}. 
All simulations finish at $t = 100 \Myr$ at the start of FG Type I SNe \citep{D'Ercole2008}. 

The FG stars are modeled by a static \cite{plummer1911} density profile: 
\begin{equation*} 
\rho_{*,\rm FG}(r) = \frac{3 \ M_{\rm FG}}{4\pi\, r_p^3} \left(1+\frac{r^2}{r_p^2}\right)^{-\frac{5}{2}},
\label{plum}
\end{equation*}
where $ r$ is the distance from the center of the cluster, $r_{\rm p}=3\pc$  and ${M_{\rm FG}}$ are the Plummer radius and the cluster mass, respectively. We neglect any changes in the FG stellar distributiofn due to cluster evolution during the simulation. We perform the simulations in the reference frame of the cluster. Therefore, we assume a static FG cluster at the center of the box and, after a time of $t_{\rm I}$, the ISM gas enters into the box from the left side with the velocity of the cluster ($23 \kms$).
Hence, the cluster is exposed to the ram pressure of the incoming gas and can accumulate it to form new stars. 

%%%%%%%%%%%%%%%%%%%%%%%%%%%%%%%%%%%%%%%%%%%%%%%%%%%%%%%%%%%%%%%%%%%%%%%%%%%%%%%%%%%%%%%%%%%%%%%%%%%
\subsection{Stellar winds of the FG stars}
Stellar winds from the AGB stars are assumed to start at $\tAGB=39\Myr$ after the formation of the FG stars. Following the models described in  \cite{calura19}, the rate of injected mass by AGB stars can be modeled as a function of time $t$ (expressed in yr)  as:
\begin{equation*}\label{AGB_ej}
\dot{\rho}_{ \rm AGB}(r)=\alpha \rho_{*,\rm FG}(r),
\end{equation*}
where $\alpha= 0.065 \, t^{-1.01}yr^{-1}$ is the specific injection rate. We add this rate as a source term into the continuity equation for gas density. We specify these ejecta with a different He abundance from the pristine gas. To do that, we define a passive scalar advected with the gas density and follow the evolution of helium (He) mass fraction for all the gas in the box.  The He mass fraction of the ejecta is assumed to slightly change from $Y(t=39\Myr)=0.36$ to $Y(t=100\Myr)= 0.32$ \citep{ventura2011}, whereas for the pristine gas we assume $Y = 0.25$. 
%%%%%%%%%%%%%%%%%%%%%%%%%%%%%%%%%%%%%%%%%%%%%%%%%%%%%%%%%%%%%%%%%%%%%%%%%%%%%%%%%%%%%%%%%%%%%
\subsection{Star formation}\label{SF}
We use the star formation model described in \cite{Rasera2006} to turn gas mass into “star” particles in eligible cells for SF based on the standard \citet{Schmidt1959} law. Thus a source term is added to the continuity equation as:
\begin{equation}
\dot{\rho}_{\rm sf}= -\frac{\rho}{t_*},
\end{equation}
where the $t_*$ is the SF timescale which is assumed to be $0.1 {\rm\, Gyr}$, as in \citet{calura19}. In this paper, we adopt four physical criteria for SF: i) a gas temperature less than $2\times10^4\K$, ii) a converging gas velocity $(\nabla\cdot \textbf{\textit{v}}<0)$,  iii) a local Jeans' length smaller than 4 (finest) cell width, i.e. less than $0.25\pc$, and iv) a density higher than $6\times10^{-23}\gocm$. When these criteria are fulfilled,  star particles are formed stochastically. The stellar particle mass is an integer multiple of ${{m}_{*} = 0.4\Msun}$, sampled from a Poisson probability distribution as described in \citet{Rasera2006}.  The star particles are placed at the center of their parent cell, with a velocity, metallicity, and He abundance equal to their natal gas. 
%%%%%%%%%%%%%%%%%%%%%%%%%%%%%%%%%%%%%%%%%%%%%%%%%%%%%%%%%%%%%%%%%%%%%%%%%%%%%%%%%%%%%%%%
\subsection{Radiative transfer}\label{sec:RT}
The \ramsesrt code is a momentum-based radiative transfer code that couples the hydrodynamics equations to the radiative transfer equations using the M1 closure for the Eddington tensor. It models the propagation of ionizing radiation and its interplay with gas via non-equilibrium thermochemistry for hydrogen and helium. The frequency range of photons is split into discrete photon groups, and then the radiative transfer equations can be solved separately for each group. In this study, we assume three photon groups (HI, HeI, and He II ionizing photons) with mean energies of 18.2, 33.0, and 61.3 eV, as described in \citetalias{Yaghoobi2022b}.  Thus in every cell, the evolution of ionization fractions for hydrogen and helium is followed \citep{rosdahl2013, Rosdahl2015}. Using this code, we study the effects of photoionization heating, cooling, and radiation pressure of ionizing radiation from the FG and SG stars as our luminous sources. Note though that additional simulations, not included in this paper, show that radiation pressure has no role in gas evolution and SF and that all effects of the ionizing radiation presented in the next sections are due to photoionization heating.
The number of injected photons by the FG and SG is estimated by a spectral energy distribution (SED) model, depending on the mass, age, and metallicity of stars. In our simulations, we use the binary population and spectral synthesis (BPASS) code \citep{BPASS2017} and assume a metallicity of $Z=0.001$ and a standard  \citet{kroupa2001} IMF for the FG stars. The same metallicity is considered for the SG stars but assuming a truncated IMF with maximum mass of $m=8\Msun$ to neglect the SN feedback effects, as assumed in \cite{calura19, Yaghoobi2022b}.  
Moreover, we assume a reduced speed of light $0.002c$ to reduce the computational cost of simulations.
%%%%%%%%%%%%%%%%%%%%%%%%%%%%%%%%%%%%%%%%%%%%%%%%%%%%%%%%%%%%%%%%%%%%%%%%%%%%%%%%%%%%%%%%%%%%%%%%%%%%%%%%%
\subsection{Cooling and Heating}\label{sec:cooling}
In \ramsesrt, radiative heating and cooling processes contribute with positive and negative rates in the energy equation, respectively. They are functions of the gas temperature, density, photon properties, and ionization fractions, as described in \cite{rosdahl2013}. The cooling processes considered for hydrogen and helium are collisional ionization, excitations, recombinations, dielectronic recombinations, bremsstrahlung, and Compton cooling, and metal cooling.

Following \citet{calura19}, we also include the heating effects of stellar winds from the AGB stars in the energy equation as:
\begin{equation}\label{EAGB}
S=0.5\alpha \rho_{*,\rm FG}\left(3\sigma^2+v^2+v_{\rm wind}^2\right),
\end{equation}
where $\sigma$ is the one-dimensional velocity dispersion of the cluster, $v_{\rm wind}$ is the wind velocity of the AGB stars, and $v$ is the local gas velocity \citep{D'Ercole2008}. We assume a wind velocity of $v_{\rm wind}=20\kms$ and an adiabatic index of $\gamma = 5/3$. The stellar winds from other sources, e. g. main-sequence stars, can be neglected due to their low mass-loss rates \citepalias{Yaghoobi2022b}. 
%----------------------------------------------------------------------------------------------
\begin{figure}
	\centering
	\includegraphics[width=1\linewidth]{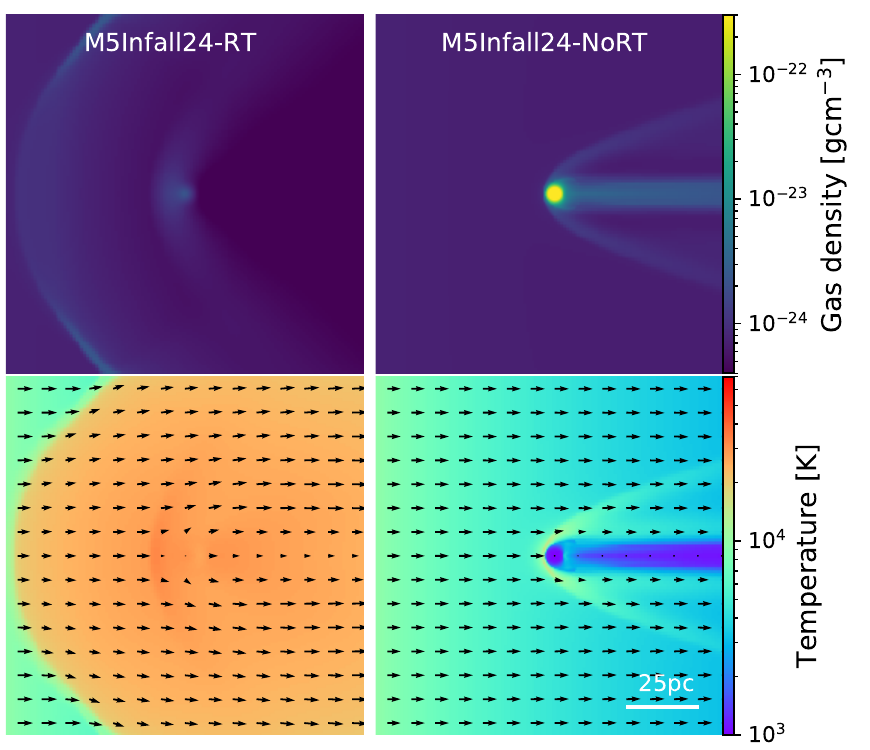}
    \caption{Gas density (top panels) and temperature (bottom panels) maps for the \MLL model in the cases with (\RT, left column) and without (\NONRT, right column) photoionization feedback at the end of the simulation, i. e. at $t = 100\Myr$ (after the FG formation). The black arrows represent the gas velocity field. Photoionization heating has caused the gas inside the cluster to expand. In contrast, radiative cooling is insufficient to cool the gas at the center of the cluster. As a result, no stars can form within $\sim 100\Myr$ in these low-mass clusters.}
 
	\label{fig:M5IN24}
\end{figure}
%-----------------------------------------------------------------------------------------------
%-----------------------------------------------------------------------------------------------
\begin{figure}
	\centering
	\includegraphics[width=1\linewidth]{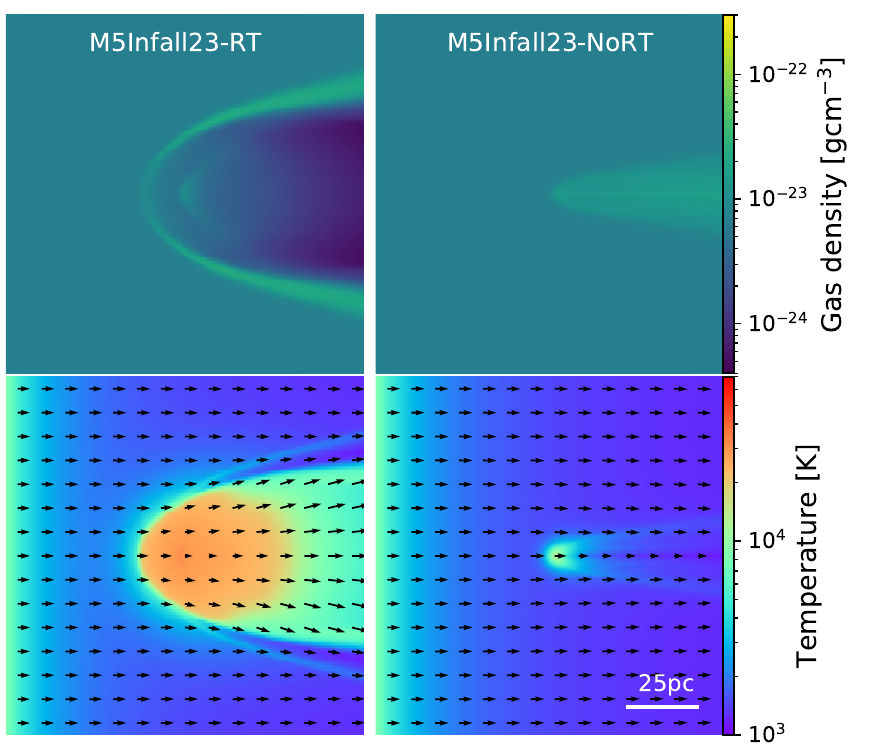}
    \caption{The same as \figref{fig:M5IN24} but for the \MLH simulation. }
    \label{fig:M5IN23}
\end{figure}
%-----------------------------------------------------------------------------------------------
%%%%%%%%%%%%%%%%%%%%%%%%%%%%%%%%%%%%%%%%%%%%%%%%%%%%%%%%%%%%%%%%%%%%%%%%%%%%%%%%%%%%%%%%%%%%%%%%%%%%%%%%%%%%
\section{Results}\label{sec:results}
Our recent studies (\citealt{calura19}; \citetalias{Yaghoobi2022a}) showed that the gravitational potential of massive clusters (${\rm 10^6\ and\ 10^7\Msun}$) can overcome ram pressure and stellar wind heating. As a result, clusters can accumulate a significant amount of gas within their central regions, where the accreted pristine gas dilutes the ejecta. In \citetalias{Yaghoobi2022b}, we showed that stellar radiative feedback can also not resist against the potential of the cluster. However, it delays the formation of SG stars in massive clusters of $10^7\Msun$ and decreases the total SG mass by $50\%$. In this section, we describe the results of our simulations listed in Table \ref{tabl1} with lower mass clusters and explore how the radiative feedback can affect the gas accumulation, of both retained AGB ejecta and accreted pristine gas.
We set our simulations to start at $ t = t_{\rm AGB}$ for \MHL and at $t = t_{\rm I}$ for three other simulations. 
Note that the $64\pc$ distance between the left side and the center of the cluster causes a time delay of about $1.8 \Myr$  for incoming gas to reach the cluster center. To take this into account, we allow the incoming gas to come into the simulation box at $t= t_{\rm I}-1.8 \Myr$  so that it passes the center at $ t_{\rm I}$. 
%%%%%%%%%%%%%%%%%%%%%%%%%%%%%%%%%%%%%%%%%%%%%%%%%%%%%%%%%%%%%%%%%%%%%%%%%%%%%%%%%%%%%%%%%%%%%%%%%%%%5
\subsection{Low-mass clusters}
In \citetalias{Yaghoobi2022a}, we found that without the effects of photoionization, the ram pressure exerted by the ambient medium can limit the ability of the gravitational potential of a $10^5\Msun$ cluster to retain its stellar winds and accrete the pristine gas. However, it did not suppress the gas accumulation at the center entirely. As a result, SG stars with high-($\approx 0.06$) and intermediate-He ($\approx0.12$) enhancements were formed within the cluster center in low- and high-density simulations, respectively. These results were in contrast with the He enhancements observed ($\leq0.01$) in low-mass clusters \citep{milone2020}. Here we include the ionizing feedback in our simulations with $10^5\Msun$ to explore how this process affects the formation of SG stars.
%-----------------------------------------------------------------------------------------------
\begin{figure*}
	\centering
	\includegraphics[width=\linewidth]{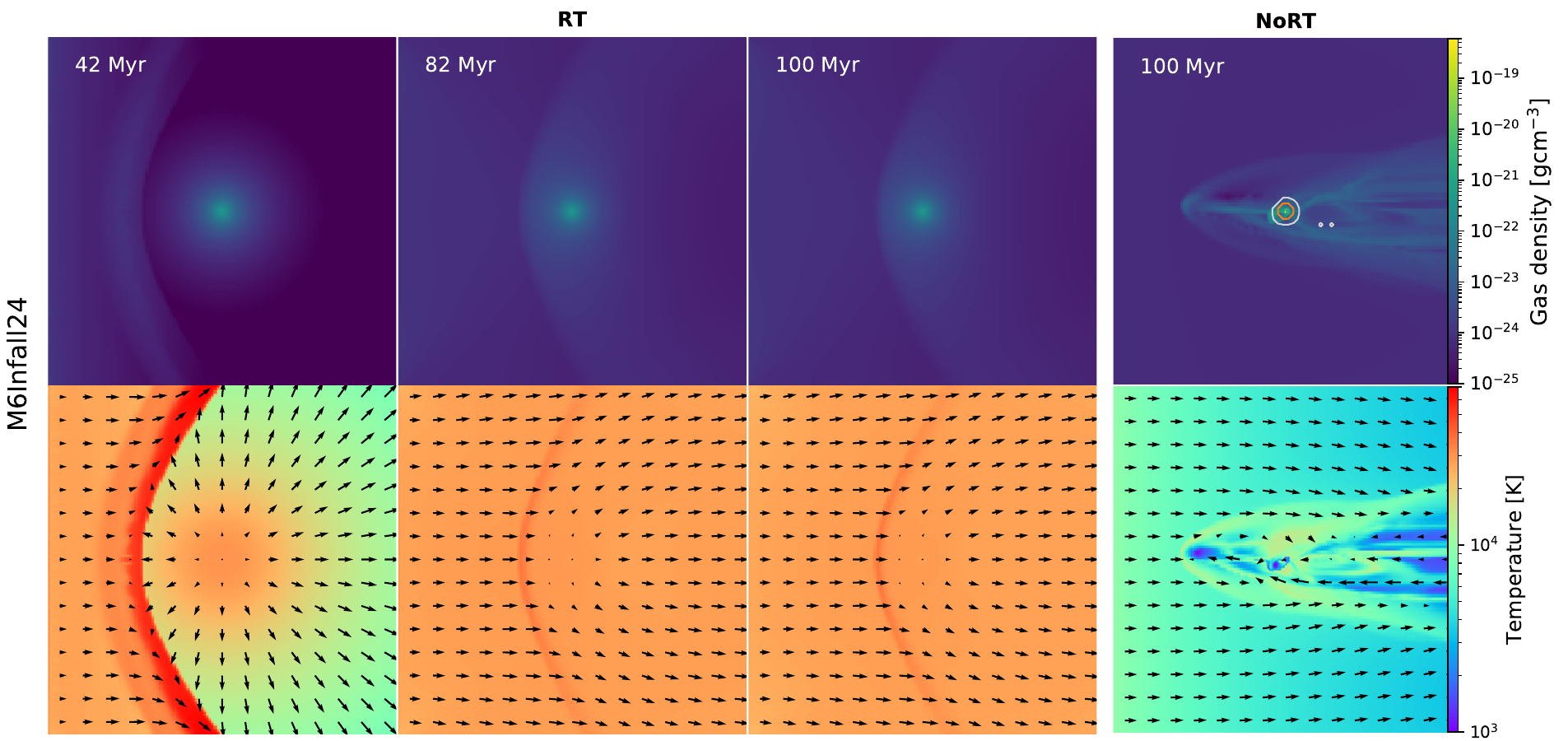}
	\includegraphics[width=\linewidth]{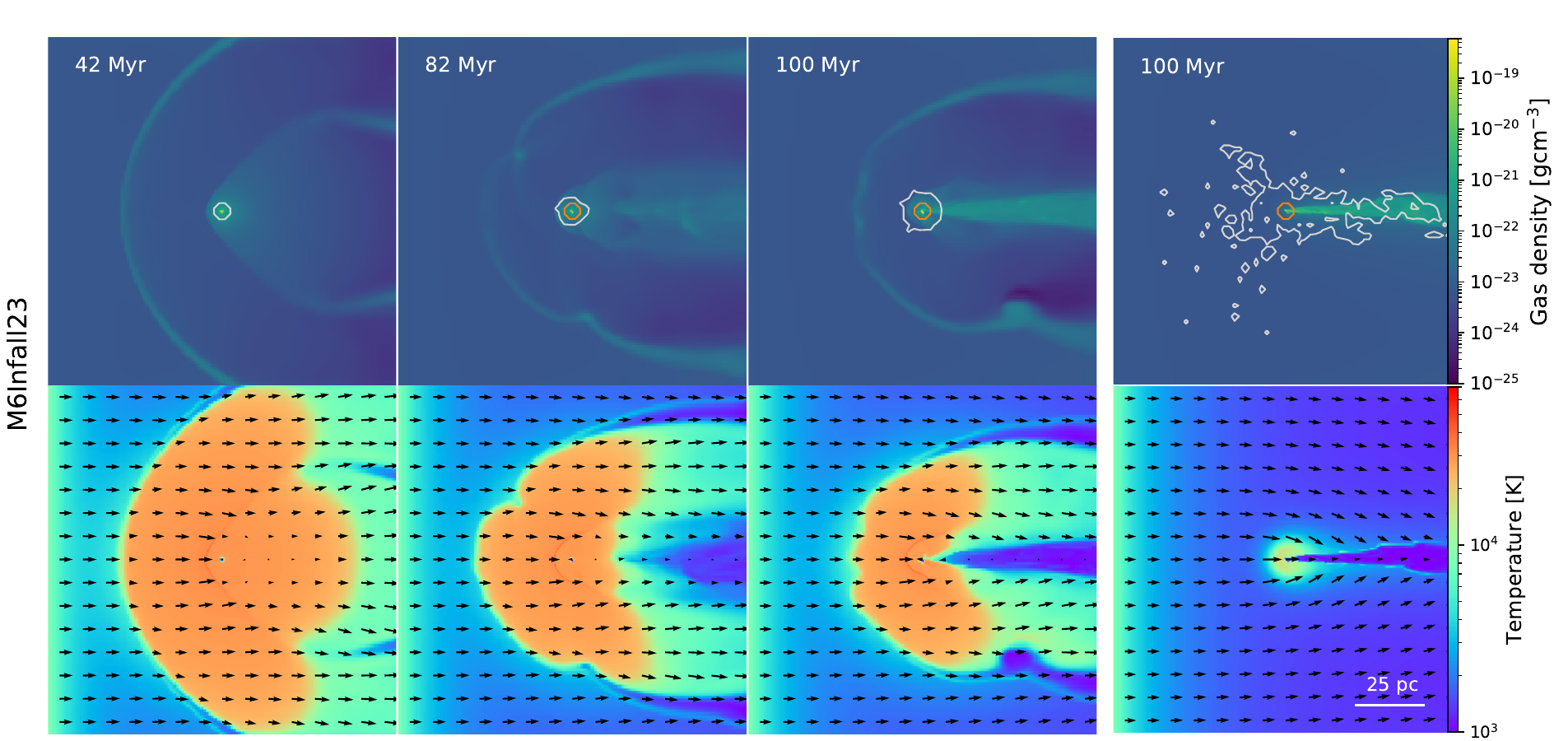}
    \caption{Gas density and temperature maps for the \MHL (first and second rows from top) and \MHH (third and fourth rows) simulations at different times for RT runs. The right column shows the results of the corresponding  \NONRT runs at the final time. 
	The white and orange contours show regions in which the SG stellar densities are $10^{-6}$ (i.e. the extent of all SG stars) and $10^{-3}$ times the maximum central density, respectively.}	
	\label{fig:M6}
\end{figure*}
%-----------------------------------------------------------------------------------------------
%%%%%%%%%%%%%%%%%%%%%%%%%%%%%%%%%%%%%%%%%%%%%%%%%%%%%%%%%%%%%%%%%%%%5
\subsubsection{\MLL}
According to the model proposed by \cite{D'Ercole2016}, we assume that the cluster in this simulation has already traversed the bubble radius created by SN feedback and reached the ISM. Therefore, the incoming gas reaches the cluster's center at $t_{\rm I} = 33.9 \Myr$ (fourth column of Table \ref{tabl1}), while the injection of mass and energy from the AGB stars begins at $\tAGB=39\Myr$.
\figref{fig:M5IN24} shows the density (upper panels) and temperature (lower panels) maps in the x-y plane for our \RT and \NONRT simulations with $10^5\Msun$ clusters at the final time ($100\Myr$). The arrows on the temperature maps indicate the velocity field of the gas. In the \NONRT simulation (right column of \figref{fig:M5IN24}, also discussed in \citetalias{Yaghoobi2022a}), the cluster could retain some of the ejecta in the central regions, accrete a marginal amount of pristine gas, and results in SG stars with a total mass of about $3\times 10^{3}\Msun$ formed from almost-pure AGB ejecta.
The left panels show that the photo-ionizing feedback of FG stars has increased the temperature and pressure of the gas within the cluster, leading to its expansion in the central regions. In the upper left panel of Figure, a shell is seen at a distance of about $60\pc$ from the center, due to the expanding bubble hitting the incoming pristine gas.
The bottom-left panel shows that the gas inside the shell is warm, with $T>10^4 K$, and therefore ionized. Hence, this shell specifies the region affected by the FG radiation around the cluster.
Moreover, the maps display a weak shock in the central regions due to the ICM hitting the ISM. Therefore, gas heating exceeds the gas cooling within the cluster, resulting in no SG star formation in this simulation. 
%%%%%%%%%%%%%%%%%%%%%%%%%%%%%%%%%%%%%%%%%%%%%%%%%%%%%%%%%%%%%%%%%%%%%55
\subsubsection{\MLH}
In the \MLH simulation, the ambient gas density is ten times more than in the previous simulation. The model starts with the infall of pristine gas at $t_{\rm I} = 30\Myr$, and about $9\Myr$ later is followed by the injection of the AGB ejecta. $t_{\rm I}$ is shorter than in the previous model because the first-generation SN feedback is expected to be less effective in pushing the ICM out into the higher-density medium. As can be seen in the right panels of \figref{fig:M5IN23}, a denser medium makes ram pressure stronger in the corresponding \NONRT case and suppresses SF but does not completely stop the accumulation of the ejecta into the cluster. However, we reported in \citetalias{Yaghoobi2022a} that a low-mass SG can be formed in the case of high-density simulation.
This different result comes from the fact that in this study, we have assumed the Jeans’ length threshold ($0.25\pc$) that prevents forming stars in cells in which the Jeans’ length is larger than four cell width (see \secref{SF}). 

The left panels of \figref{fig:M5IN23} correspond to the gas density and temperature maps of the \RT simulation of \MLH at the final time. Again, two shocks are visible in the central regions but not close to the left boundary. The region affected by the ionizing heating is less extended than in the previous simulation due to a denser infall. Within this region, the gas density is much lower than in the other parts of the box 
due to the higher pressure and temperature of the gas inside the shock. Therefore, the gas conditions inside the cluster are again not appropriate for forming SG stars within the first $100\Myr$ from the formation of the FG stars. 

Thus, in low-mass clusters, photo-ionization heating and ram pressure overcome the FG potential, prevent a significant accumulation of cold gas inside the cluster, and result in no SG star formation within $100 \Myr$ of the FG formation.
%
%%%%%%%%%%%%%%%%%%%%%%%%%%%%%%%%%%%%%%%%%%%%%%%%%%%%%%%%%%%%%%%%%%%%%%%%%%%%%%%%%%%%%%%%%%%%%%%%%%%
\subsection{Massive clusters}
From theoretical consideration, a $10^6\Msun$ cluster can overcome the ram pressure and accumulate gas in the \NONRT case \citep{lin2007}. We numerically confirmed this conclusion in  \citetalias{Yaghoobi2022a} and showed that not only could the cluster retain its AGB ejecta and accrete the pristine gas, but also that efficient cooling occurred in the central regions, leading to the formation of a massive SG. The more massive the cluster, the more accreted pristine gas, and the more massive the SG cluster and with lower He abundances. We now study the effects of ionizing radiation on SG formation for this cluster mass. 
%%%%%%%%%%%%%%%%%%%%%%%%%%%%%%%%%%%%%%%%%%%%%%%%%%%%%%%%%%%%%%
\subsubsection{\MHL}
A massive cluster in a low-density medium is assumed to create the largest SN bubble among our models. Thus, it takes more time for the incoming gas to reach the cluster's center.  Therefore, this simulation starts with the injection of the AGB ejecta at $39 \Myrs$ after the birth of the FG stars, and after about $3.5\Myr$ the incoming pristine gas passes through the cluster center. 

\figref{fig:M6} shows the density and temperature maps at different times for our simulations with massive clusters.
The white contours on the density maps display the SG stars formed in the simulation box. 
The leftmost panel in the top row shows that the AGB ejecta are already accumulating at the center of the cluster at $t=42 \Myr$, even before the infall reaches the center. All the gas in the simulation box is immediately exposed to the FG radiation and becomes ionized. As a result, at distances larger than $\sim 10 \pc$ from the center of the cluster, the photoionization heating by FG stars pushes the ICM off, as can be seen from the arrows in the temperature map. After the cluster enters the dense ISM, all the gas within the box becomes ionized and warm. 
As the simulation proceeds, a slight gas accumulation is seen in the cluster center until the end, but it does not result in sufficient cooling for SF.  
This situation is more or less unchanged until the end of the simulation.
The minimum temperature in this run is $3\times 10^4\K$. Therefore, the suitable conditions for SF are not met in this case. The two upper right panels show the final results of the \NONRT simulation in which a SG with the mass of $\approx 6\times 10^4\Msun$ has been formed. Comparing to the final results of the \RT with \NONRT simulations, the ionizing radiation suppresses the SF at least for $100\Myr$, as predicted by \cite{Conroy2011}. 
%%%%%%%%%%%%%%%%%%%%%%%%%%%%%%%%%%%%%%%%%%%%%%%%%%%%%%%%%%%%%%%%
\subsubsection{\MHH}
In a denser medium, the infall reaches the center at $t=33.9$, and the cluster accretes more pristine gas. Eventually, efficient cooling occurs only at the center of the box and stars start forming at about $40\Myr$ after the formation of the FG. 
The two bottom left panels of \figref{fig:M6} display that two shocks formed at $t=42 \Myr$, as seen in the previous simulations.
The white contour on the density map shows the first SG stars formed at this time. Moreover, the dense shell due to the ionizing radiation is formed ahead of the cluster at a distance of about $40\pc$ from the center, showing that the radiation of FG stars is not sufficient to ionize all the gas in the simulation box, in contrast to \MHL.
As the simulation proceeds, a cold gaseous tail is formed behind the cluster due to the higher gas accretion rate. However, this tail has a lower density than in the simulation without radiation, resulting in no star formation on the tail, unlike the \NONRT case (see the last column). As a consequence, SG stars are more concentrated in the cluster center than in the \NONRT case.

\begin{figure}
\centering
\includegraphics[width=\linewidth]{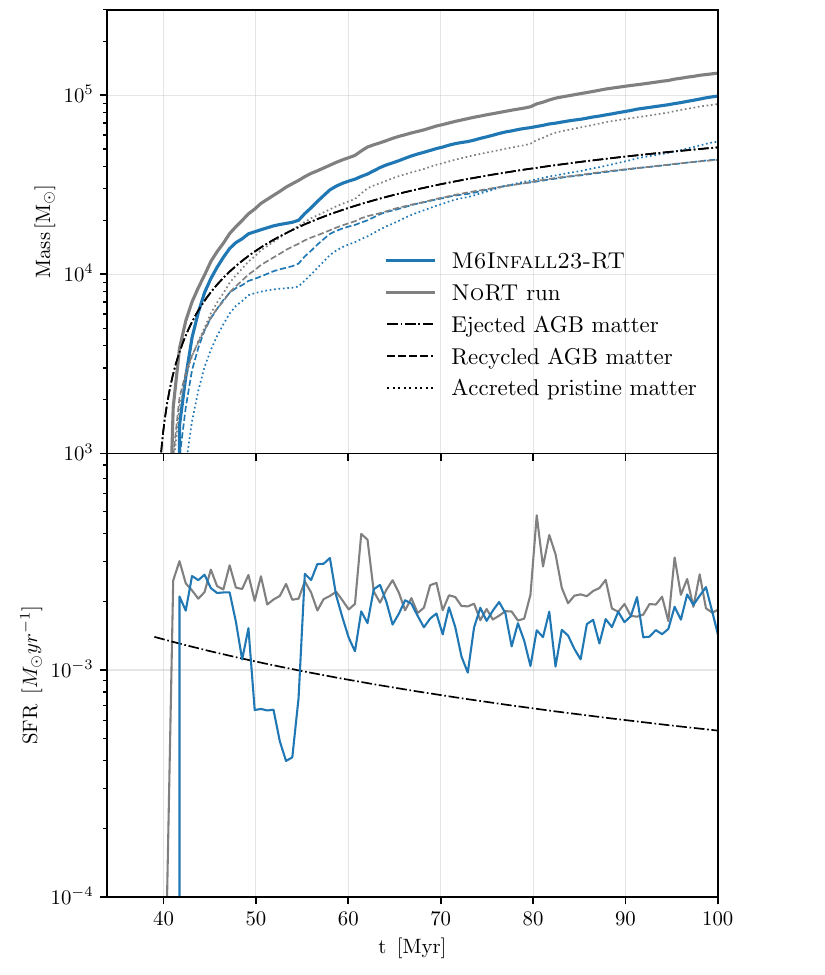}
\caption{Top: cumulative stellar mass of SG stars formed in the \MHH simulation and corresponding NoRT run as a function of time, along with the contribution from processed AGB matter and pristine gas. Bottom: SFR of the SG versus time for these runs. The lines are color-coded as follows: the solid, dashed, and dotted lines show the final SG stellar mass, the mass formed from AGB ejecta, and the mass formed from the pristine gas, respectively. The dash-dotted lines display the rate of stellar mass return injected into the simulation box.}
\label{fig:SFR}
\end{figure}

The top and bottom panels of \figref{fig:SFR} show the cumulative SG mass and the star formation rate (SFR) versus time for the \RT and \NONRT simulations, in which the mass contributions of the pristine gas (dotted line) and  AGB ejecta (dashed line) are shown. Moreover, the mass injection rate from the AGB stars is shown with dash–dotted lines. As seen in the top panel,  the amount of recycled AGB ejecta is the same for both \RT and \NONRT simulations at the end of the simulations, whereas the contribution of the accreted pristine gas in SF has decreased by almost $50\%$ in the case of \RT. 
Therefore, the inclusion of ionizing radiation leads to a $30\%$ drop in the final SG mass ($10^5\Msun$) due to the lower amount of accreted pristine gas. This result is in agreement with previous analytic studies, showing that the accreted mass decreases with higher temperature \citep{naiman2011}.
The bottom panel shows that the ionizing radiation produces a $2\Myr$ delay in star formation. In the \NONRT simulation, the SFR fluctuates around an almost constant value. This rate is larger than the rate of mass injection from AGB stars and can be attributed to the accretion rate of the pristine gas. In the case of the \RT run, the SFR experiences a visible drop at $\approx 50 \Myr$ related to the ionizing feedback of the first SG stars and then follows approximately the \NONRT rate until the end of the simulation.

\begin{figure}
\centering
\includegraphics[width=\linewidth]{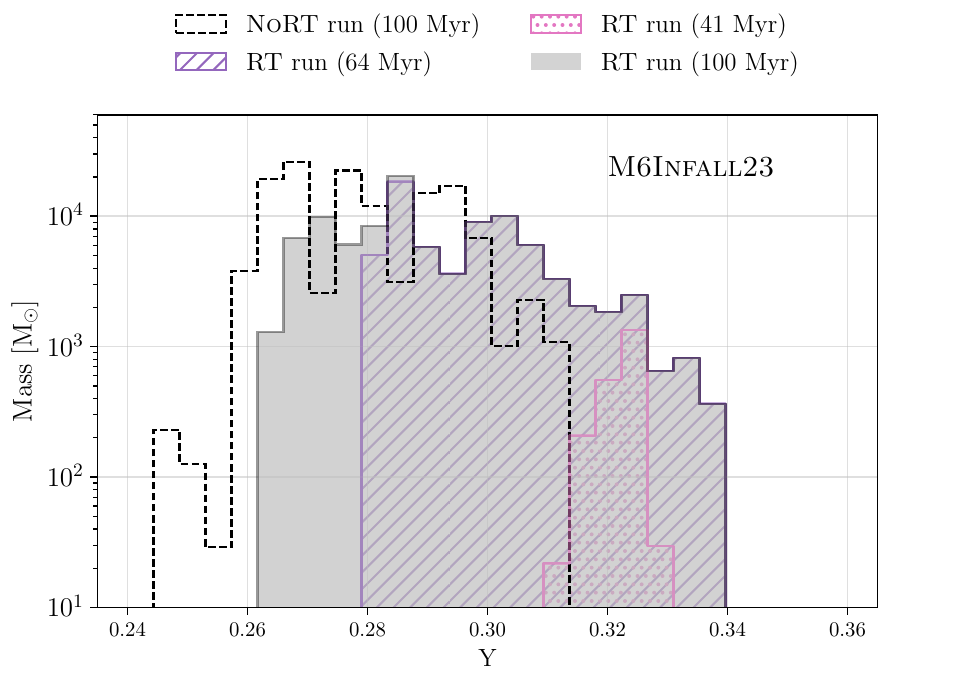}
\caption{Mass distribution of SG stars versus He abundance ${\rm Y}$ for the \MHH simulation at different times. The He mass fraction of the SG stars varies between ${\rm Y} = 0.26$, which is a little higher than the helium mass fraction of the pristine ISM gas (and FG stars, ${\rm Y} = 0.245$) at the end of the simulation, and ${\rm Y} = 0.34$, corresponding to the SG stars formed from less-diluted ejecta at the beginning of the simulations.}
\label{fig:distrib}
\end{figure}

\figref{fig:distrib} shows the He distribution of SG stars formed in the simulations at different times. The SG stars with the highest He abundances are formed in the first Myrs of the simulation when the dilution degree of the AGB ejecta is still low. Later the cluster accretes further pristine gas and the distribution eventually tends towards lower helium abundances. Comparing the final \NONRT (dashed line) and \RT distributions (gray region), SG stars generally have a higher He mass fraction in the \RT simulation because the ionizing radiation limits the pristine gas accretion. While no extremely He-rich (${\rm Y~0.36}$)  stars are formed in this run, the maximum He abundance of SG stars is $0.34$. 
%%-----------------------------------------------------------------------------------------------
%%-----------------------------------------------------------------------------------------------
%------------------------------------------------------------------------------------------------
%-----------------------------------------------------------------------------------------------
%-----------------------------------------------------------------------------------------------
\begin{table*}
	\centering
	\caption{Main results obtained at the end of \MHH simulation in this study and in \citetalias{Yaghoobi2022b} with a $10^7\Msun$ cluster. From left to right, the columns show the name of the model, the mass of the FG cluster, the final SG mass, the final SG mass fraction, the fraction of AGB ejecta, pristine gas incorporated in SG stars, and the minimum and maximum He abundance of SG stars.}  
		\begin{tabular}{lcccccccr} % four columns, alignment for each
		\hline
		Model & FG mass \ [$\Msun$] &  SG mass \ [$\Msun$] & $f_{ \rm SG/FG}$ & $f_{\rm AGB}$  &$f_{\rm P}$  &$Y_{\rm min}$   &$Y_{\rm max}$  \\
		\hline
		\MHH & $10^6$       & 1.0$\times10^5$ &0.1  & 0.9 & 0.1  &0.26 &0.34  \\
		{\sc M7Infall23}  & $10^7$  & 64.0$\times10^6$ &0.64 &0.1 & 0.9  &0.247 &0.36 \\
        {\sc M7Infall24} & $10^7$  & 6.0$\times10^5$ &0.06 &0.6 & 0.4  &0.345 &0.36  \\

		\hline
	\end{tabular}
	
	\label{tabl2}
\end{table*}
%-----------------------------------------------------------------------------------------------
%%%%%%%%%%%%%%%%%%%%%%%%%%%%%%%%%%%%%%%%%%%%%%%%%%%%%%%%%%%%%%%%%%%%%%%%%%%%%%%%%%%%%%%%%%%%%%%%%%%%%%%%%
\section{Discussion}\label{sec:discussion}
In a previous study \citepalias{Yaghoobi2022b}, we investigated the effects of ionizing feedback on the formation of SG stars in a very massive cluster ($10^7\Msun$)  in the context of the AGB scenario of \citet{D'Ercole2016}. 
To explore the role of different environments in SG star formation, we assumed that the cluster moves through a medium of different densities, i.e.  $10^{-24}$ and $10^{-23}\gocm$. In the low- and high-density simulations, we found that the total SG masses formed within $100\Myr$ after the FG formation were a fraction of $0.06$ and $0.6$ of the FG mass, respectively. In this work, we expand the study to FG clusters with the different masses of $10^5$ and $10^6\Msun$ and assume a half-mass radius of $4\pc$, estimated to be typical for proto-GCs and YMCs \citep{Baumgardt2019,Krumholz2019}. In both our low- and high-density simulations with the cluster mass $10^5\Msun$, the radiative heating dominates the ram pressure and gravitational potential of the cluster. Hence, it does not allow the gas inside the cluster to accumulate at its center, and as a result, star formation is suppressed throughout the entire time of the simulation ($100\Myr$,  \figref{fig:M5IN24} and \ref{fig:M5IN23}).
For $10^6\Msun$ clusters, our results are density-dependent. In the low-density simulation (\MHL), the FG ionizing radiation heats and expands the gas inside the cluster, resulting in no SG stars formed until the end of the simulation (the first two rows in \figref{fig:M6}). On the other hand, in the high-density case (\MHH), a higher accretion rate on the cluster occurs, causing the ICM to cool efficiently at the cluster's center. Eventually, a more compact SG cluster than in the \NONRT case is formed, with a mass of $0.1$ of FG mass (\figref{fig:M6}). 

To explore the role of ionizing feedback in the SG formation, we assume the maximum possible luminosities for our stellar populations, including binary stars (BPASS model, \citealt{BPASS2017}). As in \citetalias{Yaghoobi2022b}, we assume a bottom-heavy IMF for the SG stars so that the mass of SG stars is truncated at $8\Msun$. Hence Type II SNe for SG stars are not included in our simulations, and SG luminosities are not as high as the FG ones.

In \tabref{tabl2}, we report the final properties of the SG population obtained at the end of our \RT simulations with cluster masses $10^6$ (this work) and $10^7\Msun$ \citepalias{Yaghoobi2022b}. The SG masses in our high-density simulations exhibit an increasing trend with FG cluster mass. The reason is that more massive clusters can accrete more pristine gas, leading to the formation of more massive SG clusters. Assuming that the FG and SG stars have the same initial mass function at $100\Myr$, the SG mass fraction and SG number ratio are equivalent. Therefore, the AGB model can qualitatively explain the observed positive correlation between the SG fraction and cluster mass for the GCs hosting MSPs (see \citetalias{Yaghoobi2022a}). However, the observed SG fractions ($0.1-0.9$) are higher than the ratios computed in our simulations. This difference might come from the fact that our simulations can account only for the first $100\Myr$ of the evolution, whereas the observed systems have undergone several Gyrs of dynamical evolution. 
The SG-to-FG ratio is expected to increase during the dynamical evolution \citep{D'Ercole2008,Vesperini2021} due to the depletion of FG stars. It is found that most SG stars are retained within clusters because they are located in the central regions.
In \citetalias{Yaghoobi2022a}, we showed that assuming factors between $5$ and $20${\footnote{As for the factor 5, \cite{Larsen2012} showed that the mass of the field stars of the Fornax dwarf galaxy can at most be 4-5 times larger than that of his
globular clusters. In case the field stars have all been lost from the GCs, this result imposes an upper limit on their initial mass. However, an important quantity currently unknown is the mass of Fornax, which in the past could have been greater than today. 
The factor $20$ comes from the argument that it is possible to show that approximately  $5\%$ (i.e. $1/20$) of the first-generation mass comes out with the right composition, in terms of He abundance and p-process elements, to produce second-generation stars \citep{Renzini2013}, assuming a $100\%$ star formation efficiency.} for the ratio between the initial mass of the cluster and the present-day mass, allows us to account for the observed correlation between SG fraction and cluster mass. 
%%-----------------------------------------------------------------------------------------------
\begin{figure}
\centering
\includegraphics[width=\linewidth]{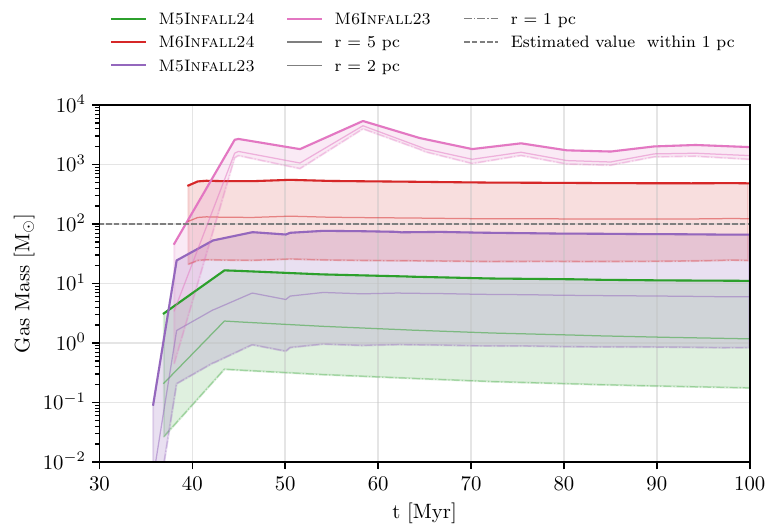}
\caption{Gas mass within different radii from $1\pc$ to $5\pc$ in the clusters studied in this paper as a function of time.}
\label{ICM}
\end{figure}
%%-----------------------------------------------------------------------------------------------
A similar conclusion can be
obtained for our RT simulations of different masses. 
The dilution of the AGB ejecta in the \RT-\MHH simulation results in the formation of SG stars with He mass fractions of $0.26<Y<0.34$. The contribution of pristine gas in forming the second generation is about 10 percent of the FG mass, which is consistent with the value proposed by \citet{D'Ercole2011}. In fact, based on the observed Na–O anticorrelation, they show that the amount of pristine gas cannot be larger than $\sim 0.1 { M_{\rm FG}}$. On the other hand, the $10^7\Msun$ clusters can form SG stars with extreme-He mass fractions  ($\approx0.36$, see last two columns of \tabref{tabl2}), showing that some of the SG stars originate out of pure AGB ejecta. It is worth noting that the dilution process of the enriched ejecta has a longer delay in massive clusters, due to the formation of a larger hot bubble by FG type II SNe around the cluster. This causes some SG stars to form from enriched materials only before the dilution of stellar winds. This result is consistent with what was proposed by \citet{D'Ercole2016} to explain the existence of extreme-He fraction in SG stars in massive GCs.
From our high-density simulations with cluster masses of $10^6$ and $10^7 \Msun$, the maximum He enhancement shows an increasing trend with cluster mass. In principle, this increasing trend can explain the observed positive correlation between maximum He enhancement and cluster mass \citep{milone2020}. The capability of reproducing this trend is a promising result for the AGB scenario.
Moreover, we found in \citetalias{Yaghoobi2022a} that in \NONRT simulations corresponding to \MLL, \MLH, and \MHL,  SG stars with extreme-He fractions could be formed, which is not in agreement with observations. This study shows that the radiative feedback suppresses the SF in these simulations and prevents the formation of these extreme-He stars. Moreover, it suppresses SF in low-mass clusters, regardless of the density of the external gas.
Therefore, our results indicate that the ionizing stellar feedback is not a severe problem for SG formation. Rather, it can help the AGB scenario to account for some observables, such as the lack of extreme He-rich stars in lowest mass GCs. However, other issues, such as the mass budget problem and the assumption of a bottom-heavy IMF for SG stars, still need further investigation.

Our study confirms that only massive GCs can form a new generation from their AGB ejecta and accreted pristine gas, as already proposed for this model \citep{D'Ercole2008,D'Ercole2016}.
Moreover, we introduce another requirement for this scenario; these massive clusters should be located in dense environments to accrete sufficient pristine gas, in order to fulfill two requirements. First, to provide a significant gas accumulation within clusters,  for the efficient cooling required for star formation. Second, to supply the pristine gas required for appropriate dilution of the AGB ejecta, so that the chemical compositions of SG stars can be comparable to observations. Therefore, our simulations with photoionization heating confirm the key role of pristine gas accretion in SG formation in diluting the AGB ejecta and can solve the abundance problem. In addition, we confirm its undeniable role in efficient cooling within young clusters.  

The density of $10^{-23}\gocm$, used in our high-density simulations, is found in star-forming disk galaxies observed at redshift $z>2$, in agreement with the \cite{D'Ercole2016} assumption that the formation of SG stars is limited to these redshifts. On the other hand, this point can explain why no strong evidence has been found so far for the presence of MSPs in YMCs \citep{Mucciarelli2008,Mucciarelli2014,Martocchia2018}. 
Probably they could not be able to accumulate a considerable amount of pristine gas at their center, because they are located in lower-density environments than the ones assumed here for the formation of MPs, and their ionizing feedback becomes effective in heating the ICM. As a result, the necessary conditions for SG star formation may not be satisfied in these clusters. This can also explain why YMCs do not show evidence of ongoing star formation and why they are gas free  \citep{Bastian2013I}. We can check the gas content within young clusters in our simulations with the caveat that the assumed metallicity in our model is lower than in local galaxies, with possible effects on radiative cooling. 

\figref{ICM} shows the evolution of the ICM mass in the radii of $1$ (dash-dotted line), $2$ (thin solid line), and $5 \pc$ (thick solid line) from the cluster’s center for the simulations of this study. To better follow how gas mass inside the cluster depends on the radius, the colored shaded regions indicate the areas covered by lines corresponding to each simulation. The gas mass quickly increases after the infall passes through the cluster center and then settles on to an approximately constant value. 
This translates into an equilibrium state between the FG gravity, ram pressure, and FG ionizing feedback. 
\citet{Bastian2013I} estimate the mass of ionized gas within a sphere of radius $1 \pc$ in YMCs according to their $H \beta$ flux. They find that the gas mass within YMCs should be lower than $100\Msun$. As can be seen in \figref{ICM}, the gas mass within $1\pc$ for those simulations in which no SG stars form (\MLL,\MLH, and \MHL), including both low-density simulations, is much lower than this value (grey dashed line), while it is larger in the \MHH simulation. 
Therefore, our study suggests that radiative feedback can prevent the accumulation of gas and star formation inside clusters of masses $\le 10^6 \Msun$ in local galaxies. 
However, further simulations specific to YMCs should be performed to confirm this result. This topic represents an interesting subject for future work.

%%%%%%%%%%%%%%%%%%%%%%%%%%%%%%%%%%%%%%%%%%%%%%%%%%%%%%%%%%%%%%%%%%%%%%%%%%%%%%%%%%%%%%%%%%%%%%%%%%%%%
\section{Conclusions}\label{Conclusions}
Using hydrodynamical simulations, we have investigated the effects of photoionization feedback on star formation and gas content in young clusters with masses of $10^5$ and $10^6\Msun$ within $30 -100\Myr$ from the birth of clusters. We assume that the clusters move in a uniform interstellar medium with gas densities of $10^{-24}$ and $10^{-23}\gocm$ to explore the effect of different environments on the amount of gas accretion onto clusters and SG formation.
These simulations are designed to check the predictions of the AGB scenario in the formation of a new generation from the stellar winds of AGB stars and accreted pristine gas in the presence of photoionization heating. 
Our main results can be summarized as follows.

\begin{itemize}

\item In the simulations with a $10^5\Msun$ cluster, the radiative heating dominates the ram pressure and gravitational potential of the cluster. Then, it does not allow the gas inside the cluster to accumulate at its center and create any low-mass stellar aggregate. As a result, star formation is suppressed until the end of simulations.

\item For $10^6\Msun$ clusters, we find that our results are
density-dependent. In the low-density simulation (\MHL), the FG ionizing radiation heats and expands the gas inside the cluster, resulting in no SG stars formed until $t=100\Myr$. On the other hand, in the high-density case (\MHH), a higher accretion rate on the cluster occurs, causing the ICM to cool efficiently at the cluster’s center. Eventually, a more compact SG cluster than in the \NONRT case is formed, with a mass of $0.1$ of FG mass. Due to ionizing effects, the cluster accretes less pristine gas, and therefore the new generation has a lower mass ($30\%$) and higher He abundances in comparison to the simulation without radiation. 

\item We confirm the key role of pristine gas accretion in SG formation and introduce another requirement for the AGB scenario: massive clusters should be located in dense environments to accrete sufficient pristine gas to fulfill two goals. First, to provide a significant gas accumulation within clusters for the efficient cooling required for star formation. Second, to supply the pristine gas required to dilute the AGB ejecta and produce an increasing trend between maximum Y variation and mass.

\item From the results of our study, it is apparent that not only the ionizing stellar feedback is not a severe problem for SG formation, but it can help the AGB scenario to account for some observables, such as the lack of extreme He-rich stars in lowest mass GCs. Moreover, the SG masses and maximum He enhancement in high-density simulations exhibit increasing trends with FG cluster mass. 

\item  In our low-density simulations, the gas content is too low so that it can be compared with estimated values for YMCs. In addition, these simulations do not show star formation, as seen in YMCs. Hence, our high-density simulations can better be matched to observations corresponding to GCs hosting MSPs, while our low-density ones can quantitatively explain the properties of young clusters in the local Universe. Therefore, the AGB model can provide a reason why YMCs do not show evidence of SG formation and any significant ICM. 

\end{itemize}

\section*{Acknowledgements}
We acknowledge support and computational resources from the PSMN (Pôle Scientifique de Modélisation Numérique) of the ENS de Lyon. AY acknowledges the support of Ferdowsi University of Mashhad.
FC acknowledges support from PRIN INAF 1.05.01.85.01, the National Recovery and Resilience Plan (Piano Nazionale di Ripresa e Resilienza, PNRR) Project ID CN 00000013 ”Italian Research Center 
on High-Performance Computing, Big Data and Quantum Computing” funded by MIUR Missione 4 Componente 2 Investimento 1.4:
Potenziamento strutture di ricerca e creazione di ”campioni nazionali di R\&S (M4C2-19)” - Next Generation EU (NGEU). 
The research activities described in this paper have been co-funded by the European Union – NextGenerationEU within PRIN 2022 project n.20229YBSAN - Globular clusters in cosmological simulations and in lensed fields: from their birth to the present epoch.

%%%%%%%%%%%%%%%%%%%%%%%%%%%%%%%%%%%%%%%%%%%%%%%%%%
\section*{Data Availability}
The data supporting this study's findings are available from the corresponding author upon reasonable request.



\bibliographystyle{mnras}
\bibliography{main} % if your bibtex file is called example.bib

\begin{thebibliography}{}
\makeatletter
\relax
\def\mn@urlcharsother{\let\do\@makeother \do\$\do\&\do\#\do\^\do\_\do\%\do\~}
\def\mn@doi{\begingroup\mn@urlcharsother \@ifnextchar [ {\mn@doi@}
  {\mn@doi@[]}}
\def\mn@doi@[#1]#2{\def\@tempa{#1}\ifx\@tempa\@empty \href
  {http://dx.doi.org/#2} {doi:#2}\else \href {http://dx.doi.org/#2} {#1}\fi
  \endgroup}
\def\mn@eprint#1#2{\mn@eprint@#1:#2::\@nil}
\def\mn@eprint@arXiv#1{\href {http://arxiv.org/abs/#1} {{\tt arXiv:#1}}}
\def\mn@eprint@dblp#1{\href {http://dblp.uni-trier.de/rec/bibtex/#1.xml}
  {dblp:#1}}
\def\mn@eprint@#1:#2:#3:#4\@nil{\def\@tempa {#1}\def\@tempb {#2}\def\@tempc
  {#3}\ifx \@tempc \@empty \let \@tempc \@tempb \let \@tempb \@tempa \fi \ifx
  \@tempb \@empty \def\@tempb {arXiv}\fi \@ifundefined
  {mn@eprint@\@tempb}{\@tempb:\@tempc}{\expandafter \expandafter \csname
  mn@eprint@\@tempb\endcsname \expandafter{\@tempc}}}

\bibitem[\protect\citeauthoryear{{Bastian} \& {Lardo}}{{Bastian} \&
  {Lardo}}{2018}]{bastian2018}
{Bastian} N.,  {Lardo} C.,  2018, \mn@doi [\araa]
  {10.1146/annurev-astro-081817-051839}, \href
  {https://ui.adsabs.harvard.edu/abs/2018ARA&A..56...83B} {56, 83}

\bibitem[\protect\citeauthoryear{{Bastian} \& {Strader}}{{Bastian} \&
  {Strader}}{2014}]{Bastian2014III}
{Bastian} N.,  {Strader} J.,  2014, \mn@doi [\mnras] {10.1093/mnras/stu1407},
  \href {https://ui.adsabs.harvard.edu/abs/2014MNRAS.443.3594B} {443, 3594}

\bibitem[\protect\citeauthoryear{{Bastian}, {Lamers}, {de Mink}, {Longmore},
  {Goodwin}  \& {Gieles}}{{Bastian} et~al.}{2013a}]{bastian2013Earlydisc}
{Bastian} N.,  {Lamers} H.~J.~G.~L.~M.,  {de Mink} S.~E.,  {Longmore} S.~N.,
  {Goodwin} S.~P.,   {Gieles} M.,  2013a, \mn@doi [\mnras]
  {10.1093/mnras/stt1745}, \href
  {https://ui.adsabs.harvard.edu/abs/2013MNRAS.436.2398B} {436, 2398}

\bibitem[\protect\citeauthoryear{{Bastian}, {Cabrera-Ziri}, {Davies}  \&
  {Larsen}}{{Bastian} et~al.}{2013b}]{Bastian2013I}
{Bastian} N.,  {Cabrera-Ziri} I.,  {Davies} B.,   {Larsen} S.~S.,  2013b,
  \mn@doi [\mnras] {10.1093/mnras/stt1779}, \href
  {https://ui.adsabs.harvard.edu/abs/2013MNRAS.436.2852B} {436, 2852}

\bibitem[\protect\citeauthoryear{{Bastian}, {Hollyhead}  \&
  {Cabrera-Ziri}}{{Bastian} et~al.}{2014}]{Bastian2014IV}
{Bastian} N.,  {Hollyhead} K.,   {Cabrera-Ziri} I.,  2014, \mn@doi [\mnras]
  {10.1093/mnras/stu1775}, \href
  {https://ui.adsabs.harvard.edu/abs/2014MNRAS.445..378B} {445, 378}

\bibitem[\protect\citeauthoryear{{Baumgardt}, {Hilker}, {Sollima}  \&
  {Bellini}}{{Baumgardt} et~al.}{2019}]{Baumgardt2019}
{Baumgardt} H.,  {Hilker} M.,  {Sollima} A.,   {Bellini} A.,  2019, \mn@doi
  [\mnras] {10.1093/mnras/sty2997}, \href
  {https://ui.adsabs.harvard.edu/abs/2019MNRAS.482.5138B} {482, 5138}

\bibitem[\protect\citeauthoryear{{Bekki}}{{Bekki}}{2017}]{Bekki2017}
{Bekki} K.,  2017, \mn@doi [\mnras] {10.1093/mnras/stx110}, \href
  {https://ui.adsabs.harvard.edu/abs/2017MNRAS.467.1857B} {467, 1857}

\bibitem[\protect\citeauthoryear{{Cabrera-Ziri} et~al.,}{{Cabrera-Ziri}
  et~al.}{2015}]{Cabrera2015V}
{Cabrera-Ziri} I.,  et~al., 2015, \mn@doi [\mnras] {10.1093/mnras/stv163},
  \href {https://ui.adsabs.harvard.edu/abs/2015MNRAS.448.2224C} {448, 2224}

\bibitem[\protect\citeauthoryear{{Cadelano}, {Dalessandro}, {Webb},
  {Vesperini}, {Lattanzio}, {Beccari}, {Gomez}  \& {Monaco}}{{Cadelano}
  et~al.}{2020}]{cadelano20}
{Cadelano} M.,  {Dalessandro} E.,  {Webb} J.~J.,  {Vesperini} E.,  {Lattanzio}
  D.,  {Beccari} G.,  {Gomez} M.,   {Monaco} L.,  2020, \mn@doi [\mnras]
  {10.1093/mnras/staa2759}, \href
  {https://ui.adsabs.harvard.edu/abs/2020MNRAS.499.2390C} {499, 2390}

\bibitem[\protect\citeauthoryear{{Calura}, {Few}, {Romano}  \&
  {D'Ercole}}{{Calura} et~al.}{2015}]{calura2015}
{Calura} F.,  {Few} C.~G.,  {Romano} D.,   {D'Ercole} A.,  2015, \mn@doi
  [\apjl] {10.1088/2041-8205/814/1/L14}, \href
  {https://ui.adsabs.harvard.edu/abs/2015ApJ...814L..14C} {814, L14}

\bibitem[\protect\citeauthoryear{{Calura}, {D'Ercole}, {Vesperini}, {Vanzella}
  \& {Sollima}}{{Calura} et~al.}{2019}]{calura19}
{Calura} F.,  {D'Ercole} A.,  {Vesperini} E.,  {Vanzella} E.,   {Sollima} A.,
  2019, \mn@doi [\mnras] {10.1093/mnras/stz2055}, \href
  {https://ui.adsabs.harvard.edu/abs/2019MNRAS.489.3269C} {489, 3269}

\bibitem[\protect\citeauthoryear{{Carretta} et~al.,}{{Carretta}
  et~al.}{2009}]{carretta2009}
{Carretta} E.,  et~al., 2009, \mn@doi [\aap] {10.1051/0004-6361/200912096},
  \href {https://ui.adsabs.harvard.edu/abs/2009A&A...505..117C} {505, 117}

\bibitem[\protect\citeauthoryear{{Chantereau}, {Biernacki}, {Martig},
  {Bastian}, {Salaris}  \& {Teyssier}}{{Chantereau}
  et~al.}{2020}]{Chantereau2020}
{Chantereau} W.,  {Biernacki} P.,  {Martig} M.,  {Bastian} N.,  {Salaris} M.,
  {Teyssier} R.,  2020, \mn@doi [\mnras] {10.1093/mnras/staa371}, \href
  {https://ui.adsabs.harvard.edu/abs/2020MNRAS.493.1306C} {493, 1306}

\bibitem[\protect\citeauthoryear{{Conroy} \& {Spergel}}{{Conroy} \&
  {Spergel}}{2011}]{Conroy2011}
{Conroy} C.,  {Spergel} D.~N.,  2011, \mn@doi [\apj]
  {10.1088/0004-637X/726/1/36}, \href
  {https://ui.adsabs.harvard.edu/abs/2011ApJ...726...36C} {726, 36}

\bibitem[\protect\citeauthoryear{{D'Ercole}, {Vesperini}, {D'Antona},
  {McMillan}  \& {Recchi}}{{D'Ercole} et~al.}{2008}]{D'Ercole2008}
{D'Ercole} A.,  {Vesperini} E.,  {D'Antona} F.,  {McMillan} S. L.~W.,
  {Recchi} S.,  2008, \mn@doi [\mnras] {10.1111/j.1365-2966.2008.13915.x},
  \href {https://ui.adsabs.harvard.edu/abs/2008MNRAS.391..825D} {391, 825}

\bibitem[\protect\citeauthoryear{{D'Ercole}, {D'Antona}  \&
  {Vesperini}}{{D'Ercole} et~al.}{2011}]{D'Ercole2011}
{D'Ercole} A.,  {D'Antona} F.,   {Vesperini} E.,  2011, \mn@doi [\mnras]
  {10.1111/j.1365-2966.2011.18776.x}, \href
  {https://ui.adsabs.harvard.edu/abs/2011MNRAS.415.1304D} {415, 1304}

\bibitem[\protect\citeauthoryear{{D'Ercole}, {D'Antona}  \&
  {Vesperini}}{{D'Ercole} et~al.}{2016}]{D'Ercole2016}
{D'Ercole} A.,  {D'Antona} F.,   {Vesperini} E.,  2016, \mn@doi [\mnras]
  {10.1093/mnras/stw1583}, \href
  {https://ui.adsabs.harvard.edu/abs/2016MNRAS.461.4088D} {461, 4088}

\bibitem[\protect\citeauthoryear{{Decressin}, {Meynet}, {Charbonnel},
  {Prantzos}  \& {Ekstr{\"o}m}}{{Decressin} et~al.}{2007a}]{decressin2007}
{Decressin} T.,  {Meynet} G.,  {Charbonnel} C.,  {Prantzos} N.,   {Ekstr{\"o}m}
  S.,  2007a, \mn@doi [\aap] {10.1051/0004-6361:20066013}, \href
  {https://ui.adsabs.harvard.edu/abs/2007A&A...464.1029D} {464, 1029}

\bibitem[\protect\citeauthoryear{{Decressin}, {Charbonnel}  \&
  {Meynet}}{{Decressin} et~al.}{2007b}]{DecCharbMey2007}
{Decressin} T.,  {Charbonnel} C.,   {Meynet} G.,  2007b, \mn@doi [\aap]
  {10.1051/0004-6361:20078425}, \href
  {https://ui.adsabs.harvard.edu/abs/2007A&A...475..859D} {475, 859}

\bibitem[\protect\citeauthoryear{{Denissenkov} \& {Hartwick}}{{Denissenkov} \&
  {Hartwick}}{2014}]{Denissenkov2014}
{Denissenkov} P.~A.,  {Hartwick} F.~D.~A.,  2014, \mn@doi [\mnras]
  {10.1093/mnrasl/slt133}, \href
  {https://ui.adsabs.harvard.edu/abs/2014MNRAS.437L..21D} {437, L21}

\bibitem[\protect\citeauthoryear{{Denissenkov}, {VandenBerg}, {Hartwick},
  {Herwig}, {Weiss}  \& {Paxton}}{{Denissenkov} et~al.}{2015}]{Denissenkov2015}
{Denissenkov} P.~A.,  {VandenBerg} D.~A.,  {Hartwick} F.~D.~A.,  {Herwig} F.,
  {Weiss} A.,   {Paxton} B.,  2015, \mn@doi [\mnras] {10.1093/mnras/stv211},
  \href {https://ui.adsabs.harvard.edu/abs/2015MNRAS.448.3314D} {448, 3314}

\bibitem[\protect\citeauthoryear{{Dickens}, {Croke}, {Cannon}  \&
  {Bell}}{{Dickens} et~al.}{1991}]{Dickens1991}
{Dickens} R.~J.,  {Croke} B.~F.~W.,  {Cannon} R.~D.,   {Bell} R.~A.,  1991,
  \mn@doi [\nat] {10.1038/351212a0}, \href
  {https://ui.adsabs.harvard.edu/abs/1991Natur.351..212D} {351, 212}

\bibitem[\protect\citeauthoryear{{Eldridge}, {Stanway}, {Xiao}, {McClelland},
  {Taylor}, {Ng}, {Greis}  \& {Bray}}{{Eldridge} et~al.}{2017}]{BPASS2017}
{Eldridge} J.~J.,  {Stanway} E.~R.,  {Xiao} L.,  {McClelland} L.~A.~S.,
  {Taylor} G.,  {Ng} M.,  {Greis} S.~M.~L.,   {Bray} J.~C.,  2017, \mn@doi
  [\pasa] {10.1017/pasa.2017.51}, \href
  {https://ui.adsabs.harvard.edu/abs/2017PASA...34...58E} {34, e058}

\bibitem[\protect\citeauthoryear{{Gavagnin}, {Bleuler}, {Rosdahl}  \&
  {Teyssier}}{{Gavagnin} et~al.}{2017}]{gavagnin2017}
{Gavagnin} E.,  {Bleuler} A.,  {Rosdahl} J.,   {Teyssier} R.,  2017, \mn@doi
  [\mnras] {10.1093/mnras/stx2222}, \href
  {https://ui.adsabs.harvard.edu/abs/2017MNRAS.472.4155G} {472, 4155}

\bibitem[\protect\citeauthoryear{{Gratton}, {Sneden}  \& {Carretta}}{{Gratton}
  et~al.}{2004}]{Gratton2004}
{Gratton} R.,  {Sneden} C.,   {Carretta} E.,  2004, \mn@doi [\araa]
  {10.1146/annurev.astro.42.053102.133945}, \href
  {https://ui.adsabs.harvard.edu/abs/2004ARA&A..42..385G} {42, 385}

\bibitem[\protect\citeauthoryear{{Gratton}, {Carretta}  \&
  {Bragaglia}}{{Gratton} et~al.}{2012}]{Gratton2012}
{Gratton} R.~G.,  {Carretta} E.,   {Bragaglia} A.,  2012, \mn@doi [\aapr]
  {10.1007/s00159-012-0050-3}, \href
  {https://ui.adsabs.harvard.edu/abs/2012A&ARv..20...50G} {20, 50}

\bibitem[\protect\citeauthoryear{{Hollyhead}, {Bastian}, {Adamo},
  {Silva-Villa}, {Dale}, {Ryon}  \& {Gazak}}{{Hollyhead}
  et~al.}{2015}]{Hollyhead2015}
{Hollyhead} K.,  {Bastian} N.,  {Adamo} A.,  {Silva-Villa} E.,  {Dale} J.,
  {Ryon} J.~E.,   {Gazak} Z.,  2015, \mn@doi [\mnras] {10.1093/mnras/stv331},
  \href {https://ui.adsabs.harvard.edu/abs/2015MNRAS.449.1106H} {449, 1106}

\bibitem[\protect\citeauthoryear{{Johnson} \& {Pilachowski}}{{Johnson} \&
  {Pilachowski}}{2010}]{Johnson2010}
{Johnson} C.~I.,  {Pilachowski} C.~A.,  2010, \mn@doi [\apj]
  {10.1088/0004-637X/722/2/1373}, \href
  {https://ui.adsabs.harvard.edu/abs/2010ApJ...722.1373J} {722, 1373}

\bibitem[\protect\citeauthoryear{{Khalaj} \& {Baumgardt}}{{Khalaj} \&
  {Baumgardt}}{2015}]{Khalaj2015}
{Khalaj} P.,  {Baumgardt} H.,  2015, \mn@doi [\mnras] {10.1093/mnras/stv1356},
  \href {https://ui.adsabs.harvard.edu/abs/2015MNRAS.452..924K} {452, 924}

\bibitem[\protect\citeauthoryear{{Krause}, {Charbonnel}, {Decressin}, {Meynet}
  \& {Prantzos}}{{Krause} et~al.}{2013}]{krause2013}
{Krause} M.,  {Charbonnel} C.,  {Decressin} T.,  {Meynet} G.,   {Prantzos} N.,
  2013, \mn@doi [\aap] {10.1051/0004-6361/201220694}, \href
  {https://ui.adsabs.harvard.edu/abs/2013A&A...552A.121K} {552, A121}

\bibitem[\protect\citeauthoryear{{Krause}, {Charbonnel}, {Bastian}  \&
  {Diehl}}{{Krause} et~al.}{2016}]{Krause2016}
{Krause} M. G.~H.,  {Charbonnel} C.,  {Bastian} N.,   {Diehl} R.,  2016,
  \mn@doi [\aap] {10.1051/0004-6361/201526685}, \href
  {https://ui.adsabs.harvard.edu/abs/2016A&A...587A..53K} {587, A53}

\bibitem[\protect\citeauthoryear{{Kravtsov} \& {Gnedin}}{{Kravtsov} \&
  {Gnedin}}{2005}]{Kravtsov2005}
{Kravtsov} A.~V.,  {Gnedin} O.~Y.,  2005, \mn@doi [\apj] {10.1086/428636},
  \href {https://ui.adsabs.harvard.edu/abs/2005ApJ...623..650K} {623, 650}

\bibitem[\protect\citeauthoryear{{Kroupa}}{{Kroupa}}{2001}]{kroupa2001}
{Kroupa} P.,  2001, \mn@doi [\mnras] {10.1046/j.1365-8711.2001.04022.x}, \href
  {https://ui.adsabs.harvard.edu/abs/2001MNRAS.322..231K} {322, 231}

\bibitem[\protect\citeauthoryear{{Kruijssen}}{{Kruijssen}}{2015}]{Kruijssen2015}
{Kruijssen} J.~M.~D.,  2015, \mn@doi [\mnras] {10.1093/mnras/stv2026}, \href
  {https://ui.adsabs.harvard.edu/abs/2015MNRAS.454.1658K} {454, 1658}

\bibitem[\protect\citeauthoryear{{Krumholz}, {McKee}  \&
  {Bland-Hawthorn}}{{Krumholz} et~al.}{2019}]{Krumholz2019}
{Krumholz} M.~R.,  {McKee} C.~F.,   {Bland-Hawthorn} J.,  2019, \mn@doi [\araa]
  {10.1146/annurev-astro-091918-104430}, \href
  {https://ui.adsabs.harvard.edu/abs/2019ARA&A..57..227K} {57, 227}

\bibitem[\protect\citeauthoryear{{Lacchin}, {Calura}  \& {Vesperini}}{{Lacchin}
  et~al.}{2021}]{lacchin21}
{Lacchin} E.,  {Calura} F.,   {Vesperini} E.,  2021, \mn@doi [\mnras]
  {10.1093/mnras/stab2061}, \href
  {https://ui.adsabs.harvard.edu/abs/2021MNRAS.506.5951L} {506, 5951}

\bibitem[\protect\citeauthoryear{{Lacchin}, {Calura}, {Vesperini}  \&
  {Mastrobuono-Battisti}}{{Lacchin} et~al.}{2022}]{Lacchin2022}
{Lacchin} E.,  {Calura} F.,  {Vesperini} E.,   {Mastrobuono-Battisti} A.,
  2022, \mn@doi [\mnras] {10.1093/mnras/stac2328}, \href
  {https://ui.adsabs.harvard.edu/abs/2022MNRAS.517.1171L} {517, 1171}

\bibitem[\protect\citeauthoryear{{Larsen}, {Strader}  \& {Brodie}}{{Larsen}
  et~al.}{2012}]{Larsen2012}
{Larsen} S.~S.,  {Strader} J.,   {Brodie} J.~P.,  2012, \mn@doi [\aap]
  {10.1051/0004-6361/201219897}, \href
  {https://ui.adsabs.harvard.edu/abs/2012A&A...544L..14L} {544, L14}

\bibitem[\protect\citeauthoryear{{Larson}}{{Larson}}{1996}]{Larson1996}
{Larson} R.~B.,  1996, in {Morrison} H.~L.,  {Sarajedini} A.,  eds,
  Astronomical Society of the Pacific Conference Series Vol. 92, Formation of
  the Galactic Halo...Inside and Out. p.~241

\bibitem[\protect\citeauthoryear{{Lin} \& {Murray}}{{Lin} \&
  {Murray}}{2007}]{lin2007}
{Lin} D. N.~C.,  {Murray} S.~D.,  2007, \mn@doi [\apj] {10.1086/515387}, \href
  {https://ui.adsabs.harvard.edu/abs/2007ApJ...661..779L} {661, 779}

\bibitem[\protect\citeauthoryear{{Marcolini}, {Brighenti}  \&
  {D'Ercole}}{{Marcolini} et~al.}{2003}]{marcolini2003}
{Marcolini} A.,  {Brighenti} F.,   {D'Ercole} A.,  2003, \mn@doi [\mnras]
  {10.1046/j.1365-2966.2003.07054.x}, \href
  {https://ui.adsabs.harvard.edu/abs/2003MNRAS.345.1329M} {345, 1329}

\bibitem[\protect\citeauthoryear{{Martocchia} et~al.,}{{Martocchia}
  et~al.}{2018}]{Martocchia2018}
{Martocchia} S.,  et~al., 2018, \mn@doi [\mnras] {10.1093/mnras/stx2556}, \href
  {https://ui.adsabs.harvard.edu/abs/2018MNRAS.473.2688M} {473, 2688}

\bibitem[\protect\citeauthoryear{{Milone} \& {Marino}}{{Milone} \&
  {Marino}}{2022}]{milone2022}
{Milone} A.~P.,  {Marino} A.~F.,  2022, \mn@doi [Universe]
  {10.3390/universe8070359}, \href
  {https://ui.adsabs.harvard.edu/abs/2022Univ....8..359M} {8, 359}

\bibitem[\protect\citeauthoryear{{Milone} et~al.,}{{Milone}
  et~al.}{2017}]{milone2017}
{Milone} A.~P.,  et~al., 2017, \mn@doi [\mnras] {10.1093/mnras/stw2531}, \href
  {https://ui.adsabs.harvard.edu/abs/2017MNRAS.464.3636M} {464, 3636}

\bibitem[\protect\citeauthoryear{{Milone} et~al.,}{{Milone}
  et~al.}{2020}]{milone2020}
{Milone} A.~P.,  et~al., 2020, \mn@doi [\mnras] {10.1093/mnras/stz2999}, \href
  {https://ui.adsabs.harvard.edu/abs/2020MNRAS.491..515M} {491, 515}

\bibitem[\protect\citeauthoryear{{Mucciarelli}, {Carretta}, {Origlia}  \&
  {Ferraro}}{{Mucciarelli} et~al.}{2008}]{Mucciarelli2008}
{Mucciarelli} A.,  {Carretta} E.,  {Origlia} L.,   {Ferraro} F.~R.,  2008,
  \mn@doi [\aj] {10.1088/0004-6256/136/1/375}, \href
  {https://ui.adsabs.harvard.edu/abs/2008AJ....136..375M} {136, 375}

\bibitem[\protect\citeauthoryear{{Mucciarelli}, {Dalessandro}, {Ferraro},
  {Origlia}  \& {Lanzoni}}{{Mucciarelli} et~al.}{2014}]{Mucciarelli2014}
{Mucciarelli} A.,  {Dalessandro} E.,  {Ferraro} F.~R.,  {Origlia} L.,
  {Lanzoni} B.,  2014, \mn@doi [\apjl] {10.1088/2041-8205/793/1/L6}, \href
  {https://ui.adsabs.harvard.edu/abs/2014ApJ...793L...6M} {793, L6}

\bibitem[\protect\citeauthoryear{{Naiman}, {Ramirez-Ruiz}  \& {Lin}}{{Naiman}
  et~al.}{2011}]{naiman2011}
{Naiman} J.~P.,  {Ramirez-Ruiz} E.,   {Lin} D.~N.~C.,  2011, \mn@doi [\apj]
  {10.1088/0004-637X/735/1/25}, \href
  {https://ui.adsabs.harvard.edu/abs/2011ApJ...735...25N} {735, 25}

\bibitem[\protect\citeauthoryear{{Naiman}, {Ramirez-Ruiz}  \& {Lin}}{{Naiman}
  et~al.}{2018}]{Naiman2018}
{Naiman} J.~P.,  {Ramirez-Ruiz} E.,   {Lin} D.~N.~C.,  2018, \mn@doi [\mnras]
  {10.1093/mnras/sty1198}, \href
  {https://ui.adsabs.harvard.edu/abs/2018MNRAS.478.2794N} {478, 2794}

\bibitem[\protect\citeauthoryear{{Piotto} et~al.,}{{Piotto}
  et~al.}{2015}]{Piotto2015}
{Piotto} G.,  et~al., 2015, \mn@doi [\aj] {10.1088/0004-6256/149/3/91}, \href
  {https://ui.adsabs.harvard.edu/abs/2015AJ....149...91P} {149, 91}

\bibitem[\protect\citeauthoryear{{Plummer}}{{Plummer}}{1911}]{plummer1911}
{Plummer} H.~C.,  1911, \mn@doi [\mnras] {10.1093/mnras/71.5.460}, \href
  {https://ui.adsabs.harvard.edu/abs/1911MNRAS..71..460P} {71, 460}

\bibitem[\protect\citeauthoryear{{Portegies Zwart}, {McMillan}  \&
  {Gieles}}{{Portegies Zwart} et~al.}{2010}]{Portegies2010}
{Portegies Zwart} S.~F.,  {McMillan} S. L.~W.,   {Gieles} M.,  2010, \mn@doi
  [\araa] {10.1146/annurev-astro-081309-130834}, \href
  {https://ui.adsabs.harvard.edu/abs/2010ARA&A..48..431P} {48, 431}

\bibitem[\protect\citeauthoryear{{Prantzos}, {Charbonnel}  \&
  {Iliadis}}{{Prantzos} et~al.}{2007}]{Prantzos2007}
{Prantzos} N.,  {Charbonnel} C.,   {Iliadis} C.,  2007, \mn@doi [\aap]
  {10.1051/0004-6361:20077205}, \href
  {https://ui.adsabs.harvard.edu/abs/2007A&A...470..179P} {470, 179}

\bibitem[\protect\citeauthoryear{{Rasera} \& {Teyssier}}{{Rasera} \&
  {Teyssier}}{2006}]{Rasera2006}
{Rasera} Y.,  {Teyssier} R.,  2006, \mn@doi [\aap]
  {10.1051/0004-6361:20053116}, \href
  {https://ui.adsabs.harvard.edu/abs/2006A&A...445....1R} {445, 1}

\bibitem[\protect\citeauthoryear{{Renzini}}{{Renzini}}{2013}]{Renzini2013}
{Renzini} A.,  2013, \mn@doi [\memsai] {10.48550/arXiv.1302.0329}, \href
  {https://ui.adsabs.harvard.edu/abs/2013MmSAI..84..162R} {84, 162}

\bibitem[\protect\citeauthoryear{{Renzini} et~al.,}{{Renzini}
  et~al.}{2015}]{renzini2015}
{Renzini} A.,  et~al., 2015, \mn@doi [\mnras] {10.1093/mnras/stv2268}, \href
  {https://ui.adsabs.harvard.edu/abs/2015MNRAS.454.4197R} {454, 4197}

\bibitem[\protect\citeauthoryear{{Rosdahl} \& {Teyssier}}{{Rosdahl} \&
  {Teyssier}}{2015}]{Rosdahl2015}
{Rosdahl} J.,  {Teyssier} R.,  2015, \mn@doi [\mnras] {10.1093/mnras/stv567},
  \href {https://ui.adsabs.harvard.edu/abs/2015MNRAS.449.4380R} {449, 4380}

\bibitem[\protect\citeauthoryear{{Rosdahl}, {Blaizot}, {Aubert}, {Stranex}  \&
  {Teyssier}}{{Rosdahl} et~al.}{2013}]{rosdahl2013}
{Rosdahl} J.,  {Blaizot} J.,  {Aubert} D.,  {Stranex} T.,   {Teyssier} R.,
  2013, \mn@doi [\mnras] {10.1093/mnras/stt1722}, \href
  {https://ui.adsabs.harvard.edu/abs/2013MNRAS.436.2188R} {436, 2188}

\bibitem[\protect\citeauthoryear{{Schmidt}}{{Schmidt}}{1959}]{Schmidt1959}
{Schmidt} M.,  1959, \mn@doi [\apj] {10.1086/146614}, \href
  {https://ui.adsabs.harvard.edu/abs/1959ApJ...129..243S} {129, 243}

\bibitem[\protect\citeauthoryear{{Teyssier}}{{Teyssier}}{2002}]{teyssier2002}
{Teyssier} R.,  2002, \mn@doi [\aap] {10.1051/0004-6361:20011817}, \href
  {https://ui.adsabs.harvard.edu/abs/2002A&A...385..337T} {385, 337}

\bibitem[\protect\citeauthoryear{{Ventura} \& {D'Antona}}{{Ventura} \&
  {D'Antona}}{2011}]{ventura2011}
{Ventura} P.,  {D'Antona} F.,  2011, \mn@doi [\mnras]
  {10.1111/j.1365-2966.2010.17651.x}, \href
  {https://ui.adsabs.harvard.edu/abs/2011MNRAS.410.2760V} {410, 2760}

\bibitem[\protect\citeauthoryear{{Vesperini}, {Hong}, {Giersz}  \&
  {Hypki}}{{Vesperini} et~al.}{2021}]{Vesperini2021}
{Vesperini} E.,  {Hong} J.,  {Giersz} M.,   {Hypki} A.,  2021, \mn@doi [\mnras]
  {10.1093/mnras/stab223}, \href
  {https://ui.adsabs.harvard.edu/abs/2021MNRAS.502.4290V} {502, 4290}

\bibitem[\protect\citeauthoryear{{Wardlow} et~al.,}{{Wardlow}
  et~al.}{2017}]{wardlow2017}
{Wardlow} J.~L.,  et~al., 2017, \mn@doi [\apj] {10.3847/1538-4357/837/1/12},
  \href {https://ui.adsabs.harvard.edu/abs/2017ApJ...837...12W} {837, 12}

\bibitem[\protect\citeauthoryear{{W{\"u}nsch}, {Palou{\v{s}}}, {Tenorio-Tagle}
  \& {Ehlerov{\'a}}}{{W{\"u}nsch} et~al.}{2017}]{Wunsch2017}
{W{\"u}nsch} R.,  {Palou{\v{s}}} J.,  {Tenorio-Tagle} G.,   {Ehlerov{\'a}} S.,
  2017, \mn@doi [\apj] {10.3847/1538-4357/835/1/60}, \href
  {https://ui.adsabs.harvard.edu/abs/2017ApJ...835...60W} {835, 60}

\bibitem[\protect\citeauthoryear{{Yaghoobi}, {Calura}, {Rosdahl}  \&
  {Haghi}}{{Yaghoobi} et~al.}{2022a}]{Yaghoobi2022a}
{Yaghoobi} A.,  {Calura} F.,  {Rosdahl} J.,   {Haghi} H.,  2022a, \mn@doi
  [\mnras] {10.1093/mnras/stab3682}, \href
  {https://ui.adsabs.harvard.edu/abs/2022MNRAS.510.4330Y} {510, 4330}

\bibitem[\protect\citeauthoryear{{Yaghoobi}, {Rosdahl}, {Calura}, {Khalaj}  \&
  {Haghi}}{{Yaghoobi} et~al.}{2022b}]{Yaghoobi2022b}
{Yaghoobi} A.,  {Rosdahl} J.,  {Calura} F.,  {Khalaj} P.,   {Haghi} H.,  2022b,
  \mn@doi [\mnras] {10.1093/mnras/stac2941}, \href
  {https://ui.adsabs.harvard.edu/abs/2022MNRAS.517.4175Y} {517, 4175}

\bibitem[\protect\citeauthoryear{{Yong}, {Grundahl}  \& {Norris}}{{Yong}
  et~al.}{2015}]{Yong2015}
{Yong} D.,  {Grundahl} F.,   {Norris} J.~E.,  2015, \mn@doi [\mnras]
  {10.1093/mnras/stu2334}, \href
  {https://ui.adsabs.harvard.edu/abs/2015MNRAS.446.3319Y} {446, 3319}

\makeatother
\end{thebibliography}

%\begin{thebibliography}{99}
%\end{thebibliography}

%%%%%%%%%%%%%%%%%%%%%%%%%%%%%%%%%%%%%%%%%%%%%%%%%%

%%%%%%%%%%%%%%%%% APPENDICES %%%%%%%%%%%%%%%%%%%%%

\appendix


%If you want to present additional material which would interrupt the flow of the main paper,it can be placed in an Appendix which appears after the list of references.

%%%%%%%%%%%%%%%%%%%%%%%%%%%%%%%%%%%%%%%%%%%%%%%%%%

% Don't change these lines
\bsp	% typesetting comment
\label{lastpage}
\end{document}